\documentclass[11pt,a4paper]{article}

\usepackage{subfigure}
\usepackage{amsthm,amsmath}
\usepackage{graphicx}
\usepackage{bm,amssymb}
\usepackage{color}
\RequirePackage[bookmarks=false,hidelinks]{hyperref}
\usepackage[font=small,labelfont=bf]{caption}
\usepackage{booktabs}
\usepackage[top=3cm, bottom=4cm, left=2.5cm, right=2.5cm]{geometry}
\usepackage{enumitem}

\newcommand{\tp}[1]{{#1^{\text{T}}}}
\DeclareMathOperator*{\logdet}{log\,det\,}
\DeclareMathOperator*{\argmax}{arg\,max\,}
\newcommand{\E}{\operatorname{E}}
\newcommand{\var}{\operatorname{Var}}
\newcommand{\cov}{\operatorname{Cov}}

\begin{document}

\title{Locally stationary spatio-temporal interpolation of \\ Argo profiling float data\medskip}
\author{Mikael Kuusela \\ \texttt{mkuusela@andrew.cmu.edu} \\ Department of Statistics and Data Science, \\ Carnegie Mellon University \and Michael L. Stein \\  \texttt{stein@galton.uchicago.edu} \\ Department of Statistics, \\ University of Chicago}
\date{}
\maketitle
\vspace{-0.3cm}

\begin{abstract}
\noindent Argo floats measure seawater temperature and salinity in the upper 2,000 m of the global ocean. Statistical analysis of the resulting spatio-temporal dataset is challenging due to its nonstationary structure and large size. We propose mapping these data using locally stationary Gaussian process regression where covariance parameter estimation and spatio-temporal prediction are carried out in a moving-window fashion. This yields computationally tractable nonstationary anomaly fields without the need to explicitly model the nonstationary covariance structure. We also investigate Student-$t$ distributed fine-scale variation as a means to account for non-Gaussian heavy tails in ocean temperature data. Cross-validation studies comparing the proposed approach with the existing state-of-the-art demonstrate clear improvements in point predictions and show that accounting for the nonstationarity and non-Gaussianity is crucial for obtaining well-calibrated uncertainties. This approach also provides data-driven local estimates of the spatial and temporal dependence scales for the global ocean which are of scientific interest in their own right.

\end{abstract}

\vspace{0.5cm}

\section{Introduction} \label{sec:intro}

The subsurface open ocean has historically been one of the least studied places on Earth, due to a lack of observational data at fine enough spatial and temporal resolutions. That changed dramatically soon after the turn of the century with the introduction of the Argo array of profiling floats. Argo is a collection of nearly 4,000 autonomous floats that measure temperature and salinity in the upper 2,000 m of the ocean. The array's nearly uniform $3^\circ \times 3^\circ \times 10$~days sampling of the global ocean has enabled oceanographers to study the subsurface ocean at unprecedented accuracy and scale. Argo data have been used, for example, to quantify global changes in ocean heat content~\cite{Roemmich2015}, to study ocean circulation \cite{Gray2014}, mesoscale eddies \cite{Zhang2014}, internal waves \cite{Hennon2014} and tropical cyclones~\cite{Cheng2015} and to improve climate model predictions \cite{Chang2013}. Argo has now become the primary source of subsurface temperature and salinity data for these and hundreds of other studies of ocean climate and dynamics.

\begin{figure}[!t]
	\centering
	\includegraphics[trim = 5cm 0cm 0cm 6cm, clip=true, width=12cm]{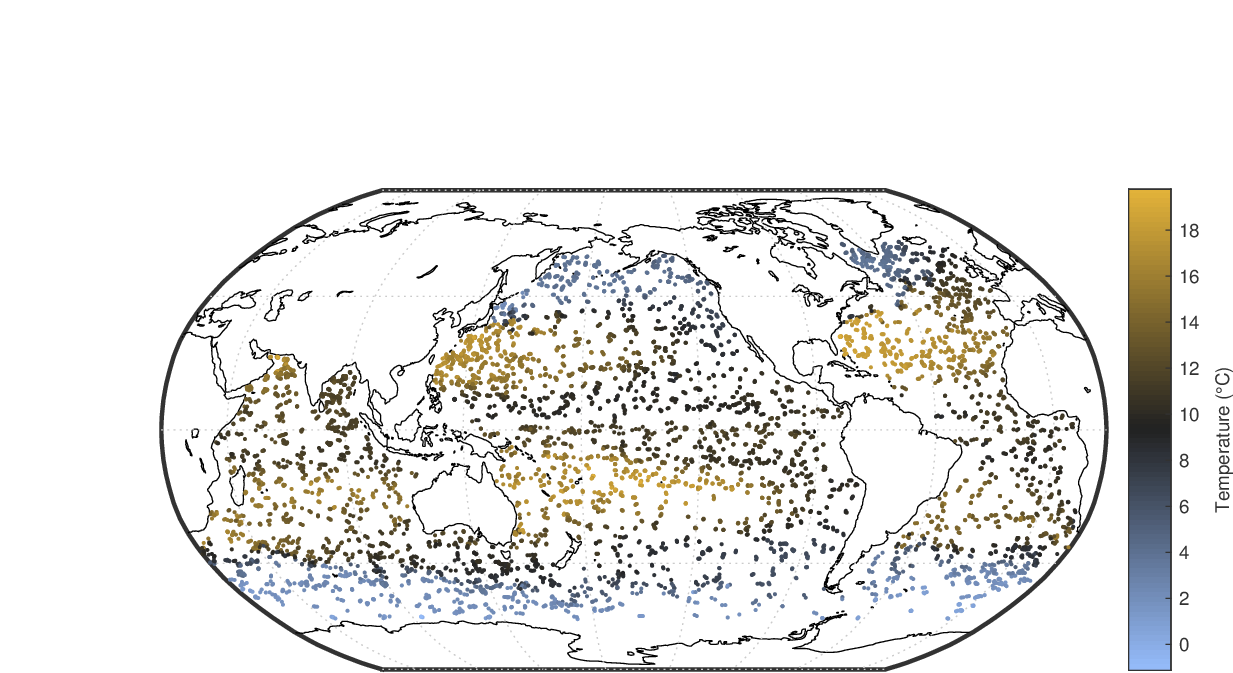}
	\caption{Argo temperature data at 300~dbar ($\approx$ 300~m) for February 2012. Argo provides \textit{in~situ} temperature and salinity observations for the upper 2,000 m of the global ocean. The statistical question studied in this work is how to interpolate these irregularly sampled data onto a dense regular grid.}
	\label{fig:interpolatedTemp300}
\end{figure}

A significant portion of scientific results from Argo rely on spatially and temporally interpolated temperature and salinity maps, such as those in \cite{Roemmich2009,Schmidtko2013,Good2013}. These data products transform the irregularly located Argo observations onto a fine regular grid which facilitates further scientific analysis. In a sense, the goal is to ``fill in the gaps'' between the \textit{in~situ} Argo observations, such as those shown in Figure~\ref{fig:interpolatedTemp300}, and to turn them into a continuously interpolated field in both space and time. To achieve this, a number of statistical modeling assumptions need to be made and the resulting interpolated maps, along with their uncertainties, may be sensitive to these choices. Given the unique nature of Argo data and the scientists' reliance on the gridded maps, it is of utmost importance to produce the interpolations using the most appropriate statistical techniques and to rigorously understand their performance and limitations.

From a statistical perspective, Argo observations constitute a fascinatingly rich geostatistical dataset. There is a huge volume of data (nearly 2 million profiles, each having between 50--1,000 vertical observations, have been collected to date), the data are nonstationary in both their mean and covariance structure and exhibit heavy tails and other non-Gaussian features. Argo is also a rare example of a truly four-dimensional (3 $\times$ space + time) \textit{in~situ} observational dataset. Standard approaches to parameter estimation and interpolation for spatio-temporal random fields do not easily scale up to datasets of this size since the number of required computations grows cubically with the number of observations. Furthermore, in order to capture the full complexity of these data, there is a need to develop models that go beyond the usual assumptions of stationarity and Gaussianity.

In this paper, we propose a statistical framework for interpolating Argo data that is aimed at addressing both the computational issues caused by the size of the dataset and the modeling challenges caused by the nonstationarity of these data. The approach is based on the basic idea that if a prediction is desired at the spatio-temporal location $(\bm{x}^*,t^*)$, where $\bm{x}^* = \tp{[x_\mathrm{lat}^*,x_\mathrm{lon}^*]}$ is the spatial location in degrees of latitude and longitude and $t^*$ is time, then the observations that are close to $(\bm{x}^*,t^*)$ in the $(\bm{x},t)$ space should be the most informative for making the prediction. (Following standard statistical terminology, we use here and throughout the term \emph{prediction} to refer to the interpolations of the unknown values of the random field.) Furthermore, while it is clear from a simple exploratory inspection of Argo data that a global random field model for the ocean would need to be nonstationary, it is reasonable to assume that the data can be \emph{locally} modeled as a stationary random field. (Assuming that the mean has been successfully removed, a spatio-temporal random field $f(\bm{x},t)$ is said to be \emph{stationary} if the covariance $k(\bm{x}_1,t_1,\bm{x}_2,t_2) = \cov(f(\bm{x}_1,t_1),f(\bm{x}_2,t_2))$ is a function of $(\bm{x}_1-\bm{x}_2,t_1-t_2)$ only, that is, $k(\bm{x}_1,t_1,\bm{x}_2,t_2) = \linebreak k(\bm{x}_1-\bm{x}_2,t_1-t_2)$. In other words, the covariance depends only on the difference of the two spatio-temporal locations and not on where in the ocean these locations are. When that is not the case, the field is said to be \emph{nonstationary}.) We combine these two ideas by considering only data within a small spatio-temporal neighborhood of $(\bm{x}^*,t^*)$ and by assuming that this subset of data can be modeled as a stationary random field. We use the data within the neighborhood to first estimate the unknown parameters of the random field model using maximum likelihood and then to perform the interpolation at $(\bm{x}^*,t^*)$. As proposed by Haas in \cite{Haas1990,Haas1995}, we use the neighborhoods in a moving-window fashion: when moving to the next grid point, the window is re-centered and the parameters re-estimated. This leads to data-driven, spatially varying estimates of the spatio-temporal dependence structure and provides gridded interpolations that reflect the nonstationarity of the underlying random field. This approach is computationally efficient since it considers only a subset of the full dataset at a time and since the computations across the grid points can be fully parallelized. In a line of work described in \cite{Hammerling2012a, Hammerling2012b, Tadic2015, Tadic2017}, a closely related moving-window method has been developed and successfully applied to mapping remote sensing data.

The Roemmich--Gilson climatology (along with the associated anomalies) \cite{Roemmich2009} is one of the more popular gridded Argo data products. Roemmich and Gilson first estimate the mean field using a weighted local regression fit to several years of Argo data and then perform kriging \cite{Cressie1993,Stein1999} (also known as optimal interpolation \cite{Daley1991} or objective analysis/mapping~\cite{Bretherton1976} and closely related to Gaussian process regression \cite{Rasmussen2006}) on the mean-subtracted monthly residuals to obtain the interpolated anomaly fields. In this work, we use the Roemmich--Gilson mean field, but improve the modeling of the anomalies in three important ways: first, we include time in the interpolation; second, we use data-driven local estimates of the nonstationary covariance structure as described above; and third, we consider Student-$t$ distributed fine-scale variation (the so-called \emph{nugget effect}) in order to account for non-Gaussian heavy tails in the data (that is, observations whose magnitude is larger than what a Gaussian distribution would produce). We investigate the point prediction and uncertainty quantification performance of the proposed approach using cross-validation studies. We demonstrate that adding the temporal component to the mapping leads to major performance improvements, while the locally estimated covariance parameters and the Student-$t$ distributed nugget effect are crucial for obtaining reasonable uncertainties. The uncertainty quantification part is particularly important since the Roemmich--Gilson data product does not currently provide any uncertainty information, presumably due to challenges in modeling the nonstationary and non-Gaussian features of the data.

It is worth highlighting that one of the key differences between this work and most previous Argo maps (see Sections~\ref{sec:background}\ref{sec:RG_climatology} and \ref{sec:background}\ref{sec:otherProducts} for a literature review) is our use of Argo-based data-driven estimates of the covariance function parameters. Using the same data to both estimate the covariance parameters and to perform the mapping is standard in the relevant statistics literature \cite{Stein1999,Rasmussen2006,Cressie1993,Cressie2011}, but so far the oceanographic community has largely not implemented these ideas (a notable exception is \cite{Gray2014,Gray2015} where ocean velocity fields are mapped based on data-driven variogram fits to Argo data). Indeed, in many Argo data products, the procedure for choosing the covariance parameters can be best described as making an informed guess. The covariance structure is typically motivated by what is known about ocean dynamics on a qualitative level. For example, the spatial length scales are typically increased near the Equator. But the quantitative details, such as which specific scales are used and how much they are varied, are typically decided in an \emph{ad hoc} manner, with limited justification. In some cases, some of the covariance length scales are set equal to the $3^\circ$ Argo sampling resolution \cite{Gaillard2012,Gaillard2009}. This seems statistically incorrect since the length scales should reflect the dependence structure of the underlying physical field and not the sampling resolution of the observing system. In this work, we avoid this kind of arbitrariness by fitting the covariance parameters with maximum likelihood to several years of Argo data itself. The fit is done using local moving windows, as described above. The estimated covariance parameters are shown to exhibit physically reasonable spatial patterns and can be of scientific interest in their own right, as they contain information about ocean dynamics in the different regions sampled by Argo. The data-driven covariance estimates are shown to improve both the point predictions and the uncertainties in comparison to the Roemmich--Gilson covariance, with the improvement in uncertainty quantification being particularly substantial.

In the present work, we focus primarily on interpolating Argo temperature anomalies, but similar techniques can also be developed for the salinity fields. The rest of this paper is structured as follows: Section~\ref{sec:background} provides an overview of the existing Argo data products with an emphasis on the Roemmich--Gilson climatology. Section~\ref{sec:methods} describes the proposed locally stationary spatio-temporal interpolation method. Section~\ref{sec:results} applies the new approach to Argo data and studies its performance in terms of point predictions, uncertainty quantification and the estimated model parameters. Section~\ref{sec:discussion} concludes and discusses directions for future~work. The supplement~\cite{Kuusela2017Supplement} provides further information and results. Readers who are unfamiliar with Argo are recommended to consult Section~1 in the supplement for an overview of Argo floats and data.

\section{Overview of previous Argo data products} \label{sec:background}

\subsection{Roemmich--Gilson climatology} \label{sec:RG_climatology}

The Roemmich--Gilson (RG) climatology and its anomalies \cite{Roemmich2009,RGWebsite2016} are constructed by first estimating a seasonally varying mean field and then performing kriging on the mean-subtracted monthly residuals. The vertical dimension is handled by binning the profiles into 58 pressure bins, whose sizes increase with depth. Each binned value is calculated so that it represents an average over the pressure bin (John Gilson, personal communication, 2017). The mean field is estimated using 3 adjacent pressure levels, but kriging for the anomalies is carried out using data from only one pressure level at a time.

The RG mean field is a weighted local least-squares spatio-temporal regression fit that is carried out separately for each latitude-longitude grid point and each pressure level. The local regression function is (John Gilson, personal communication, 2016)
\begin{align}
	m(x_\mathrm{lat},x_\mathrm{lon},z,t) = \beta_0 &+ [\,\text{first- and second-order terms of } x_\mathrm{lat}, x_\mathrm{lon} \text{ and } z\,] \notag \\ &+ \sum_{k=1}^6 \gamma_k \sin\left(2\pi k \frac{t}{365.25}\right) + \sum_{k=1}^6 \delta_k \cos\left(2\pi k \frac{t}{365.25}\right), \label{eq:RGregFun}
\end{align}
where $x_\mathrm{lat}$ is latitude, $x_\mathrm{lon}$ is longitude, $z$ is pressure and $t$ is time in yeardays. This function is fitted to $3 \times 12 \times 100$ nearest neighbors, where the factors refer to the 3 pressure levels and 12 calendar months. The nearest neighbors are found across the entire Argo dataset and are given weights according to their horizontal distance from the grid point. The horizontal distance metric~is
\begin{equation}
	d_\mathrm{RG}(\bm{x}_1,\bm{x}_2) = \sqrt{(\Delta x_\mathrm{lat})^2 + (\Delta x_\mathrm{lon})^2 + \mathrm{Pen}(\bm{x}_1,\bm{x}_2)^2}, \label{eq:RGdist}
\end{equation}
where $\bm{x}_i = \tp{[x_{\mathrm{lat},i},x_{\mathrm{lon},i}]}$, $\Delta x_\mathrm{lat}$ is the meridional distance in kilometers, $\Delta x_\mathrm{lon}$ is the zonal distance in kilometers (converted from degrees to kilometers using the midpoint latitude) and $\mathrm{Pen}(\bm{x}_1,\bm{x}_2)$ is a penalty term for crossing ocean depth contours (see \cite{Roemmich2009} for details). Once fitted, the regression function is evaluated at the grid point for the midpoint of each month to produce the monthly mean field estimates for that latitude, longitude and pressure.

The anomalies are computed one month at a time. Here the crucial modeling choice concerns the covariance structure of those fields. The Roemmich--Gilson covariance function is
\begin{equation}
	k(\bm{x}_1,\bm{x}_2) \propto 0.77 \exp\left(-\left(\frac{d_{\mathrm{RG},a}(\bm{x}_1,\bm{x}_2)}{140\text{ km}}\right)^2\right) + 0.23 \exp\left(-\frac{d_{\mathrm{RG},a}(\bm{x}_1,\bm{x}_2)}{1111\text{ km}}\right), \label{eq:RGcovariance}
\end{equation}
where $d_{\mathrm{RG},a}(\bm{x}_1,\bm{x}_2)$ is otherwise the same distance metric as in Equation~\eqref{eq:RGdist}, but with $\Delta x_\mathrm{lon}$ replaced by $a\big((x_{\mathrm{lat},1}+x_{\mathrm{lat},2})/2\big) \cdot \Delta x_\mathrm{lon}$, where \cite{Roemmich2009,RGWebsite2016}
\begin{equation}
	a(x_\mathrm{lat}) = \begin{cases}
		1, \quad \text{if } |x_\mathrm{lat}| > 20^\circ, \\
		\frac{1}{8} + \frac{7}{160^\circ} |x_\mathrm{lat}|, \quad \text{if } |x_\mathrm{lat}| \leq 20^\circ.
	\end{cases}
\end{equation}
Using $d_{\mathrm{RG},a}(\bm{x}_1,\bm{x}_2)$ instead of $d_\mathrm{RG}(\bm{x}_1,\bm{x}_2)$ has the effect of elongating the zonal dependence in the tropics. The result is a nonstationary covariance function that varies in the zonal direction depending on the latitude. However, the meridional range is kept constant and the same covariance function is used at all longitudes, all pressure levels, all seasons and for both the temperature and the salinity anomalies. The noise-to-signal variance ratio (the ratio of the nugget variance and the Gaussian process variance $\sigma^2/\phi$ in the terminology and notation of Section~\ref{sec:methods}) is set to be 0.15 throughout the global ocean.

The functional form and the parameter values in Equation~\eqref{eq:RGcovariance} are motivated by the observed empirical correlation for the steric height anomaly (essentially a vertical integral of the density anomaly; see Section~7.6.2 in \cite{Talley2011}) in Argo and satellite altimetry data (see Figure~2.2 in \cite{Roemmich2009}). The parameter values are chosen by a graphical comparison of the correlation functions instead of a formal estimation procedure, such as maximum likelihood or weighted least squares. Due to atmospheric interaction, the length scales are expected to be longer in the mixed layer than in the deeper ocean. Since the RG covariance is chosen based on a vertically integrated quantity, it is likely that the length scales chosen this way are too short near the surface and too long at larger~depths.

An important limitation of the Roemmich--Gilson climatology is that it does not provide uncertainty estimates. In principle, the formal kriging variance could be used to provide Gaussian prediction intervals, but this would require defining the proportionality constant in Equation~\eqref{eq:RGcovariance} (this constant cancels out in the kriging point predictions). Even if this proportionality constant was provided, the uncertainties are unlikely to be reliable given that the RG covariance does not vary with pressure and uses a fixed noise-to-signal variance ratio. Another limitation is that the RG covariance is only a function of the spatial locations and does not take the temporal dimension into account.

Gasparin et al. \cite{Gasparin2015} improve the Roemmich--Gilson model by including time in the covariance and by adding nonstationarity in the meridional range parameter. They also briefly investigate the prediction uncertainties. But their analysis is limited to the Equatorial Pacific only, their covariance remains the same for all pressure levels and longitudes and their covariance parameters are not based on formal statistical estimates. In this work, we go further and develop data-driven covariances for Argo temperature data that can vary as a function of latitude, longitude, pressure and season. Covariance estimates, anomaly maps and uncertainties are investigated in the global ocean at three exemplary pressure levels. The covariances include time and nonstationarity is allowed in all covariance parameters, including the zonal, meridional and temporal range parameters.

\subsection{Other data products and analysis techniques} \label{sec:otherProducts}

Besides the Roemmich--Gilson climatology, several other Argo-based data products have been produced. We give here a brief overview of some of these products and the underlying statistical methods. Our treatment is by no means exhaustive---a full list of gridded Argo data products can be found at \url{http://www.argo.ucsd.edu/Gridded_fields.html}.

One of the defining features of the MIMOC product \cite{Schmidtko2013} is that it is mapped on isopycnals (surfaces of constant density) instead of pressure surfaces. This approach may have distinct advantages in handling the vertical movement of water masses and in avoiding density inversions. Statistically, temperature and salinity maps on isopycnals are likely to be less nonstationary than on pressure surfaces. However, in order to enable conversion from density coordinates to pressure coordinates, one needs to provide maps of pressure on the isopycnals and these fields remain highly nonstationary. MIMOC also incorporates techniques for improving the mapping in areas of sharp fronts and varying bathymetry. The EN4 data product \cite{Good2013} provides uncertainty estimates and includes the vertical dimension in the covariance model, but its uncertainty quantification procedure seems statistically \emph{ad hoc}, is only validated in the root-mean-square sense and shows signs of miscalibration below roughly 400~m. ISAS \cite{Gaillard2012} and MOAA~GPV~\cite{Hosoda2008} are further examples of kriging-based Argo data products. For most products, there is some effort to use physical data to justify the chosen covariance parameters but no formal statistical estimators of these parameters are used. An exception is the work of Gray and Riser \cite{Gray2014,Gray2015} which uses a weighted least-squares variogram fit and an iteratively estimated mean field to map ocean velocity fields based on Argo data.

The above-mentioned products are all variants of kriging-based spatial or spatio-temporal interpolation. However, various other techniques are also available for analyzing oceanographic data. LOESS regression has proved useful for estimating the mean field and the seasonal cycle from Argo~\cite{Ridgway2002}. This is the basis for the Roemmich--Gilson mean field as well as for the CARS2009 data product \cite{CARS2009}. However, kriging is still needed for obtaining the monthly anomalies. Model-driven data assimilation is commonly used for assimilating Argo data to ocean reanalysis products, examples include ORAS5 \cite{Zuo2018}, ECCO \cite{Forget2015} and GODAS \cite{Behringer2004}. Reanalysis products may, however, have non-negligible biases due to the assumed dynamical model and the intricacies of the data assimilation algorithm. Empirical orthogonal functions (EOFs; also known as principal component analysis to statisticians) are also often used in analyzing oceanographic data and especially satellite observations. Since computing the EOFs requires repeated observations at the same spatial locations, ungridded Argo data cannot directly be used for deriving them. An alternative would be to define the leading EOFs using a simulation model followed by a fit to Argo data, but this would make the analysis dependent on the quality of the simulation and would be likely to oversmooth small-scale features.

\section{Locally stationary interpolation of Argo data} \label{sec:methods}

This section describes the statistical methodology we propose for interpolating Argo temperature data. The approach is based on a locally stationary spatio-temporal Gaussian process (GP) regression model. Here a \emph{Gaussian process} refers to a random function, whose values at any finite set of locations follow a multivariate Gaussian distribution, see, e.g., Chapter~2 in \cite{Rasmussen2006}. We start by linearly interpolating the Argo temperature profiles to a given pressure level. We then subtract the seasonally varying Roemmich--Gilson mean field (see Sections~\ref{sec:background}\ref{sec:RG_climatology} and \ref{sec:results}\ref{sec:setupAndData}) and work with the residuals, which are assumed to have zero mean. Our goal is to use these residuals to produce gridded maps of temperature anomalies. Similar to the Roemmich--Gilson anomalies, our analysis is carried out separately for each pressure level.

Let $(\bm{x}^*,t^*)$ with $\bm{x}^* = \tp{[x_\mathrm{lat}^*,x_\mathrm{lon}^*]}$ be a space-time grid point for which a prediction is desired. We assume that within a small spatio-temporal neighborhood $\mathcal{W}(\bm{x}^*,t^*) = [x^*_\mathrm{lat}-x_\mathrm{win},x^*_\mathrm{lat}+x_\mathrm{win}] \times [x^*_\mathrm{lon}-x_\mathrm{win},x^*_\mathrm{lon}+x_\mathrm{win}] \times [t^*-t_\mathrm{win},t^*+t_\mathrm{win}]$ around $(\bm{x}^*,t^*)$ the following model~holds:
\begin{equation}
	y_{i,j} = f_i(\bm{x}_{i,j},t_{i,j}) + \varepsilon_{i,j}, \quad f_i \overset{\mathrm{iid}}{\sim} \mathrm{GP}(0,k(\bm{x}_1,t_1,\bm{x}_2,t_2;\bm{\theta})), \label{eq:spatioTemporalModel}
\end{equation}
where $i=1,\ldots,n$ refers to years and $j=1,\ldots,m_i$ to observations within $\mathcal{W}(\bm{x}^*,t^*)$ at the desired pressure level in the $i$th year, $y_{i,j}$ is the $(i,j)$th mean-subtracted temperature residual, $\bm{x}_{i,j} = \tp{[x_{\mathrm{lat},i,j},x_{\mathrm{lon},i,j}]}$ and $t_{i,j}$ are the location (in degrees of latitude and longitude) and time (in yeardays) of $y_{i,j}$ and $\mathrm{GP}(0,k(\bm{x}_1,t_1,\bm{x}_2,t_2;\bm{\theta}))$ denotes a zero-mean Gaussian process with a stationary space-time covariance function $k(\bm{x}_1,t_1,\bm{x}_2,t_2;\bm{\theta}) = k(\bm{x}_1{-}\bm{x}_2,t_1{-}t_2;\bm{\theta})$ depending on parameters~$\bm{\theta}$. We use an anisotropic exponential space-time covariance function $k(\bm{x}_1,t_1,\bm{x}_2,t_2;\bm{\theta}) = \phi \exp\left( -d(\bm{x}_1,t_1,\bm{x}_2,t_2)\right)$, where the GP variance $\phi > 0$,
\begin{equation}
	d(\bm{x}_1,t_1,\bm{x}_2,t_2) = \sqrt{\left( \frac{x_{\mathrm{lat},1} - x_{\mathrm{lat},2}}{\theta_\mathrm{lat}} \right)^2 + \left( \frac{x_{\mathrm{lon},1} - x_{\mathrm{lon},2}}{\theta_\mathrm{lon}} \right)^2 + \left( \frac{t_1 - t_2}{\theta_t} \right)^2} \label{eq:distanceMetric}
\end{equation}
and $\theta_\mathrm{lat}$, $\theta_\mathrm{lon}$ and $\theta_t$ are positive range parameters. The term $\varepsilon_{i,j}$ in \eqref{eq:spatioTemporalModel} is called the \emph{nugget effect} and is included in the model to capture fine-scale variation. It is independent of $f_i$ and assumed to follow either $\varepsilon_{i,j} \overset{\mathrm{iid}}{\sim} N(0,\sigma^2)$, where $N(0,\sigma^2)$ is the zero-mean Gaussian distribution with variance $\sigma^2$, or $\varepsilon_{i,j} \overset{\mathrm{iid}}{\sim} t_\nu(\sigma^2)$, where $t_\nu(\sigma^2)$ is the scaled Student-$t$ distribution with $\nu > 1$ degrees of freedom and scale parameter $\sigma > 0$ (that is, $Z \sim t_\nu(\sigma^2)$ if and only if $\frac{Z}{\sigma} \sim t_\nu$, where $t_\nu$ is the Student-$t$ distribution with $\nu$ degrees of freedom). The Student nugget provides a way to model non-Gaussian heavy tails in the observed data. Notice that under this model we obtain $n$ independent realizations of the random field, one for each year, which facilitates estimating the model parameters $(\bm{\theta},\sigma^2)$ or $(\bm{\theta},\sigma^2,\nu)$.

\begin{figure}[!t]
	\centering
	\includegraphics[trim = 0cm 0cm 0cm 0cm, clip=true, width=7cm]{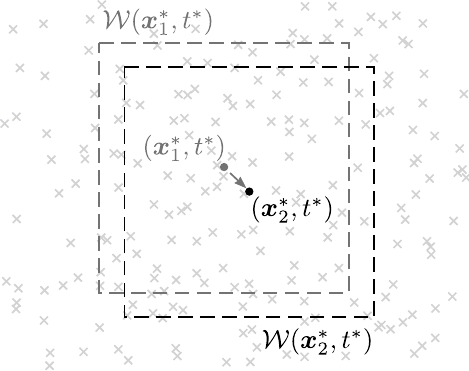}
	\caption{Illustration of moving-window Gaussian process regression. In order to make a prediction at the grid point $(\bm{x}_1^*,t^*)$, data in its local neighborhood $\mathcal{W}(\bm{x}_1^*,t^*)$ are used to first estimate the covariance parameters and then to make the prediction. When moving to the next grid point $(\bm{x}_2^*,t^*)$, the window moves along and the parameter estimates and the prediction are made using data within the new local neighborhood $\mathcal{W}(\bm{x}_2^*,t^*)$. For ease of graphical presentation, the time axis is suppressed here, but the same concept applies in the full spatio-temporal space.}
	\label{fig:movingWindowGP}
\end{figure}

We employ model \eqref{eq:spatioTemporalModel} in a moving-window fashion: for each grid point $(\bm{x}^*,t^*)$, we use data within the local neighborhood $\mathcal{W}(\bm{x}^*,t^*)$ to estimate the model parameters and to predict $y_i^* = f_i(\bm{x}^*,t^*) + \varepsilon_i^*$. When we move to the next grid point, we re-center the window around the new location and re-estimate the model parameters. The overlap of the nearby windows results in smoothly varying local estimates of the model parameters. Such a moving-window approach to Gaussian process regression was proposed by Haas in \cite{Haas1990,Haas1995}. Figure~\ref{fig:movingWindowGP} illustrates the method.

This approach facilitates both the modeling and the computational challenges in Argo data analysis. Instead of having to specify a global nonstationary covariance model, the moving-window approach enables us to handle the nonstationarity in Argo data using a collection of locally fitted stationary GP models. In terms of the computations, the approach essentially replaces the inversion of one large covariance matrix by many inversions of smaller covariance matrices. This results in significant computational gains, especially since the computations across the grid points are embarrassingly parallel. More concretely, let $n$ be the number of years and $m$ the typical number of global Argo observations per year. Then, to leading order, the computing time of a global GP regression fit would be $n \cdot C \cdot m^3$, where $C$ is a constant. When each moving window contains fraction $f$ of data, there are $N$ grid points and the computations are parallelized to $p$ threads, the computing time of the moving-window approach is $\frac{1}{p} \cdot N \cdot n \cdot C \cdot (fm)^3 = \frac{1}{p} \cdot N \cdot f^3 \cdot n \cdot C \cdot m^3$. Rough values for the analysis carried out in this paper are $f = 0.01$, $N = 30{,}000$ and $p = 30$, leading to a speed-up factor $\big(\frac{1}{p} \cdot N \cdot f^3\big)^{-1} = 1{,}000$.

In model \eqref{eq:spatioTemporalModel}, we understand the nugget effect $\varepsilon_{i,j}$ to primarily reflect ocean variability at spatio-temporal scales smaller than the $3^\circ \times 3^\circ \times 10$~days Argo sampling resolution. The fitted nugget may also capture sensor noise, but we consider this component to be negligibly small in comparison to fine-scale ocean variability. This interpretation of the nugget leads us to make predictions for $y_i^* = f_i(\bm{x}^*,t^*) + \varepsilon_i^*$ instead of $f_i(\bm{x}^*,t^*)$, which widens the prediction intervals by an amount corresponding to the fine-scale component $\varepsilon_i^*$. This is a sensible approach for temperature data, since fine-scale ocean temperature variability is orders of magnitude larger than the noise-level of the Argo temperature sensors. For salinity, especially at large depths, it would be appropriate to model the measurement process more carefully, but this is outside the scope of the present work.

Some previous works \cite{Gasparin2015,Gray2015,Gaillard2012,Gaillard2009,Schmidtko2013} on space-time modeling of Argo data use \emph{separable} covariance models in which the covariance factorizes into a function of space and a function of time. Such covariance models do not allow the dependence in space to interact with the dependence in time (see, e.g., Section 6.1.3 in \cite{Cressie2011}) and hence imply that large-scale spatial features decay in time just as quickly as small-scale features, which is unrealistic for most real-world processes. The model~\eqref{eq:distanceMetric} is in contrast \emph{nonseparable} and implies that large-scale spatial features decay more slowly than small-scale features, as one would expect.

To demonstrate the importance of proper modeling of temporal effects, we also consider a spatial version of model~\eqref{eq:spatioTemporalModel}. In that case, the moving window remains the same as before, but we ignore the temporal separation of the observations within the window by setting $\theta_t = \infty$. This is equivalent to dropping the time covariate $t_{i,j}$ in Equation~\eqref{eq:spatioTemporalModel}. The next two sections explain in more detail the parameter estimation, point prediction and uncertainty quantification steps under model \eqref{eq:spatioTemporalModel}.

\subsection{Gaussian nugget} \label{sec:calculationsGaussianNugget}

With the Gaussian nugget $\varepsilon_{i,j} \sim N(0,\sigma^2)$, the unknown model parameters are the covariance parameters $\bm{\theta} = \tp{[\phi, \theta_\mathrm{lat}, \theta_\mathrm{lon}, \theta_t]}$ and the nugget variance $\sigma^2$. We use maximum likelihood to estimate these parameters from Argo data. For each year $i=1,\ldots,n$, let $\bm{K}_i(\bm{\theta})$ be the $m_i \times m_i$ matrix with elements $\big[\bm{K}_i(\bm{\theta})\big]_{j,k} = k(\bm{x}_{i,j},t_{i,j},\bm{x}_{i,k},t_{i,k};\bm{\theta})$ and let $\bm{y}_i$ be the column vector with elements $y_{i,j},\,j=\nobreak1,\ldots,m_i$. The log-likelihood of the parameters $(\bm{\theta},\sigma^2)$ is
\begin{align}
	\ell(\bm{\theta},\sigma^2) &= \sum_{i=1}^n \log p(\bm{y}_i|\bm{\theta},\sigma^2) \\ &= -\frac{1}{2} \sum_{i=1}^n \Big[ \logdet(\bm{K}_i(\bm{\theta}) + \sigma^2 \bm{I}) + \tp{\bm{y}_i} (\bm{K}_i(\bm{\theta}) + \sigma^2\bm{I})^{-1} \bm{y}_i + m_i \log(2\pi) \Big]
\end{align}
and the maximum likelihood estimator (MLE) of $(\bm{\theta},\sigma^2)$ is
\begin{equation}
	(\bm{\hat{\theta}},\hat{\sigma}^2) = \argmax_{(\bm{\theta},\sigma^2) \in \mathbb{R}_+^5} \ell(\bm{\theta},\sigma^2).
\end{equation}
The MLE needs to be obtained using numerical optimization; we use the BFGS quasi-Newton algorithm, as implemented in the Matlab Optimization Toolbox~R2016a \cite{Mathworks2016}, on log-transformed parameters.

We base the predictions on the conditional distribution $p(y_i^*|\bm{y}_i,\bm{\hat{\theta}},\hat{\sigma}^2)$, where we have plugged in the MLE $(\bm{\hat{\theta}},\hat{\sigma}^2)$ for the model parameters. Standard manipulations show that the predictive distribution is
\begin{equation}
	(y_i^*|\bm{y}_i,\bm{\hat{\theta}},\hat{\sigma}^2) \sim N\left(\tp{(\bm{k}_i^*(\bm{\hat{\theta}}))}(\bm{K}_i(\bm{\hat{\theta}})+\hat{\sigma}^2 \bm{I})^{-1}\bm{y}_i,\, \hat{\phi} + \hat{\sigma}^2 - \tp{(\bm{k}_i^*(\bm{\hat{\theta}}))}(\bm{K}_i(\bm{\hat{\theta}})+\hat{\sigma}^2\bm{I})^{-1} \bm{k}_i^*(\bm{\hat{\theta}})\right), \label{eq:predictiveDistGaussNug}
\end{equation}
where $\bm{k}_i^*(\bm{\theta})$ is a column vector with elements $k(\bm{x}^*,t^*,\bm{x}_{i,j},t_{i,j};\bm{\theta}),\,j=1,\ldots,m_i$. We make the point predictions using the conditional mean $\hat{y}_i^* = \E(y_i^*|\bm{y}_i,\bm{\hat{\theta}},\hat{\sigma}^2) = \tp{(\bm{k}_i^*(\bm{\hat{\theta}}))}(\bm{K}_i(\bm{\hat{\theta}})+\hat{\sigma}^2 \bm{I})^{-1}\bm{y}_i$. This is the well-known kriging predictor (see, e.g., \cite{Cressie1993,Stein1999}) based on the data within the moving window $\mathcal{W}(\bm{x}^*,t^*)$ and with plug-in values for the model parameters. It is mean square error optimal assuming that model \eqref{eq:spatioTemporalModel} is correct and ignoring the uncertainty of the model parameters. To quantify our uncertainty about $y_i^*$, we use the $1-\alpha$ predictive intervals based on~$p(y_i^*|\bm{y}_i,\bm{\hat{\theta}},\hat{\sigma}^2)$,
\begin{equation}
	\big[\,\underline{y}_i^*,\,\overline{y}_i^*\,\big] = \left[\,\hat{y}_i^* - z_{1-\alpha/2} \sqrt{\var(y_i^*|\bm{y}_i,\bm{\hat{\theta}},\hat{\sigma}^2)},\, \hat{y}_i^* + z_{1-\alpha/2} \sqrt{\var(y_i^*|\bm{y}_i,\bm{\hat{\theta}},\hat{\sigma}^2)}\,\right],
\end{equation}
where $z_{1-\alpha/2}$ is the $1-\alpha/2$ standard normal quantile and $\var(y_i^*|\bm{y}_i,\bm{\hat{\theta}},\hat{\sigma}^2) = \hat{\phi} + \hat{\sigma}^2 - \linebreak \tp{(\bm{k}_i^*(\bm{\hat{\theta}}))}(\bm{K}_i(\bm{\hat{\theta}})+\hat{\sigma}^2\bm{I})^{-1} \bm{k}_i^*(\bm{\hat{\theta}})$ is the kriging variance.

\subsection{Student nugget} \label{sec:calculationsStudentNugget}

With the Student nugget $\varepsilon_{i,j} \sim t_\nu(\sigma^2)$, the likelihood $p(\bm{y}|\bm{\theta},\sigma^2,\nu) = \prod_{i=1}^n p(\bm{y}_i|\bm{\theta},\sigma^2,\nu)$ and the predictive distribution $p(y_i^*|\bm{y}_i,\bm{\theta},\sigma^2,\nu)$ are not available in closed form. To achieve computationally tractable inferences, we employ the Laplace approximation as in \cite{Vanhatalo2009}; see also Section 3.4 in \cite{Rasmussen2006}. The Laplace approximation yields tractable likelihood computations for estimating the unknown model parameters, including the degrees of freedom $\nu$. It also provides closed-form approximate point predictions, while the prediction intervals can be obtained using Monte Carlo sampling. The computations were implemented by adapting the \texttt{GPML} toolbox~\cite{Rasmussen2010,Rasmussen2016}. Further details are provided in Section~2 in the supplement \cite{Kuusela2017Supplement}.

\section{Results} \label{sec:results}

In this section, we study the performance of the statistical methodology described in Section~\ref{sec:methods} in interpolating Argo temperature data. We use the Roemmich--Gilson approach described in Section~\ref{sec:background}\ref{sec:RG_climatology} as a baseline and investigate how the data-driven local estimates of the covariance structure and the inclusion of time improve the interpolation of the anomalies. Section~\ref{sec:results}\ref{sec:setupAndData} describes the models and datasets that we consider. We then study the fitted models from three perspectives: 1) cross-validated point prediction performance in Section~\ref{sec:results}\ref{sec:pointPred}, 2) cross-validated uncertainty quantification performance in Section~\ref{sec:results}\ref{sec:UQ} and 3) the estimated spatio-temporal dependence structure in Section~\ref{sec:results}\ref{sec:localDependence}.

\subsection{Experiment setup} \label{sec:setupAndData}

Our experiments are performed by fitting models to Argo temperature data collected between 2007 and 2016. We investigate three pressure levels, 10, 300 and 1500~decibars (dbar), with the model parameters estimated separately for each pressure level. We focus on 1- or 3-month temporal windows centered around February (i.e., February 2007, February 2008,$\,\ldots\,$, February~2016 or January--March 2007, January--March 2008,$\,\ldots\,$, January--March 2016 are considered independent realizations from the statistical model in Equation~\eqref{eq:spatioTemporalModel}). To enable comparison of models with different time windows, all the cross-validation studies are done for February data~only.

We consider six different statistical models (Table~\ref{tab:models}). Model~1, which we regard as the baseline reference model, is our reimplementation of the Roemmich--Gilson maps~\cite{Roemmich2009}; see Section~\ref{sec:background}\ref{sec:RG_climatology}. Apart from a few technical details (see below), this model is the same as the one used in~\cite{Roemmich2009}. Models~2--6 are variants of the locally stationary mapping procedure described in Section~\ref{sec:methods}. In each case, the mapping is carried out on a $1^\circ \times 1^\circ$ grid and the covariance parameters are estimated using maximum likelihood within a $20^\circ \times 20^\circ$ moving window on this grid. Model~2 is a 1-month fit with a purely spatial covariance $k(\bm{x}_1,\bm{x}_2;\bm{\theta})$ and a Gaussian nugget. Model~3 is otherwise the same but with a Student nugget. Model~4 is a 3-month version of Model~2. Models~5~and~6 are 3-month fits with a spatio-temporal covariance $k(\bm{x}_1,t_1,\bm{x}_2,t_2;\bm{\theta})$ and either a Gaussian or a Student nugget. The spatio-temporal fits are done with 3 months of data to make sure that there are enough profiles from each float to estimate the temporal covariance structure (there are usually 9 profiles from each float within a 3-month window).

For each model, the mean field is the Roemmich--Gilson local regression fit (see Section~\ref{sec:background}\ref{sec:RG_climatology}). The publicly available version of the RG climatology \cite{RGWebsite2016} includes only the annual mean field, but John Gilson kindly provided us with the mid-month evaluations of the local mean functions~\eqref{eq:RGregFun}. We use either these mid-month evaluations treating the mean as a constant over the temporal window (spatial mean) or alternatively a temporally varying reconstruction of the original mean field (spatio-temporal mean) as indicated in Table~\ref{tab:models}. As a function of time, the local regression function \eqref{eq:RGregFun} has 13 free parameters which we wish to reconstruct from the 12 mid-month evaluations. We do this by using the Moore--Penrose pseudoinverse which handles the one extra degree of freedom by finding the minimum-norm solution in the space of the regression coefficients \cite{Harville2008}.

\begin{table}[t]
	\caption{Description of the models we consider for interpolating Argo temperature data. Model~1 is a reimplementation of the procedure developed by Roemmich and Gilson (RG) in \cite{Roemmich2009} and Models 2--6 are variants of locally stationary interpolation. The models differ in terms of how time is taken into account, in the distribution of the fine-scale variation represented by the nugget effect and in the length of the temporal window used in the fit (see text for more details).}
	\label{tab:models}
	\centering
	\begin{tabular}{lllll}
		\toprule
		Model & Time window & Mean & Covariance & Nugget \\
		\midrule
		1 & February & RG (spatial) & RG-like & Gaussian \\
		2 & February & RG (spatial) & Local (spatial) & Gaussian \\
		3 & February & RG (spatial) & Local (spatial) & Student \\
		4 & January--March & RG (spatio-temporal) & Local (spatial) & Gaussian \\
		5 & January--March & RG (spatio-temporal) & Local (spatio-temporal) & Gaussian \\
		6 & January--March & RG (spatio-temporal) & Local (spatio-temporal) & Student \\
		\bottomrule 
	\end{tabular}
\end{table}

The reference model (Model~1) implements the key aspects of the Roemmich--Gilson approach. There are, however, two technical differences between our implementation and theirs. First, we do not include the depth penalty term $\mathrm{Pen}(\bm{x}_1,\bm{x}_2)$ when computing the point estimates and the uncertainties. And, second, we do not implement any specialized treatment of distances across islands or continental land. These differences are unlikely to markedly affect the conclusions drawn below. Finally, we note that when computing the prediction intervals for the reference model, we need to provide the proportionality constant in Equation~\eqref{eq:RGcovariance} (i.e., the variance $\phi$ of the Gaussian process part of the spatial model). Since Roemmich and Gilson do not provide uncertainty estimates, they also do not give estimates of this constant as it cancels out in the point predictions. To simulate what potentially could have been done to produce uncertainties under the Roemmich--Gilson model, we estimate the proportionality constant using a moving-window empirical variance. That is, the RG-inspired prediction intervals at the grid point $\bm{x}^*$ are formed using the covariance model~\eqref{eq:RGcovariance} with the following estimate of the GP variance $\phi$:
\begin{equation}
	\hat{\phi} = \frac{\text{empirical variance of } y_{i,j} \text{ in a } 20^\circ \times 20^\circ \times 1 \text{ month window centered at } \bm{x}^*}{1.15},
\end{equation}
where the denominator originates from the RG noise-to-signal variance ratio 0.15 via the relation $\var(y_{i,j}) = \phi + \sigma^2 = \phi \left( 1 + \frac{\sigma^2}{\phi} \right) = 1.15 \cdot \phi$. We emphasize that the resulting prediction intervals are not part of the original Roemmich--Gilson climatology and are by no means advocated by them for uncertainty quantification.

We fit Models 1--6 to the global Argo dataset as of May 8, 2017 \cite{ArgoData2017}. The quality control criteria used for filtering out profiles with technical issues are given in the supplement \cite{Kuusela2017Supplement}. There were a total of 1,417,813 Argo profiles in 2007--2016, out of which 994,709 passed our selection criteria. We also performed a further temporal filtering to focus on the desired time windows and a spatial filtering based on the Roemmich--Gilson land mask \cite{RGWebsite2016}, which filters out profiles located in marginal seas, such as the Mediterranean Sea or the Gulf of Mexico. The final dataset has 70,227 profiles for February and 223,797 profiles for January to March.

The analysis was carried out using Matlab R2016a. As noted in Section~\ref{sec:methods}\ref{sec:calculationsStudentNugget}, we use \texttt{GPML} \cite{Rasmussen2010,Rasmussen2016} to fit Models 3 and 6, while the other models are our own implementations. The computations were carried out on the Midway2 cluster at the University of Chicago Research Computing Center. The possibility to parallelize the moving-window computations to the 28~threads of the Midway2 compute nodes was crucial for making the analysis computationally feasible. The Matlab code used to produce the results is available on Github \cite{Kuusela2017Github}.

\subsection{Point predictions} \label{sec:pointPred}

\begin{figure}[t]
	\centering
	
	\subfigure[10 dbar (local space-time)]{
		\includegraphics[width=7.7cm,trim = 5cm 0cm 0cm 6cm, clip=true]{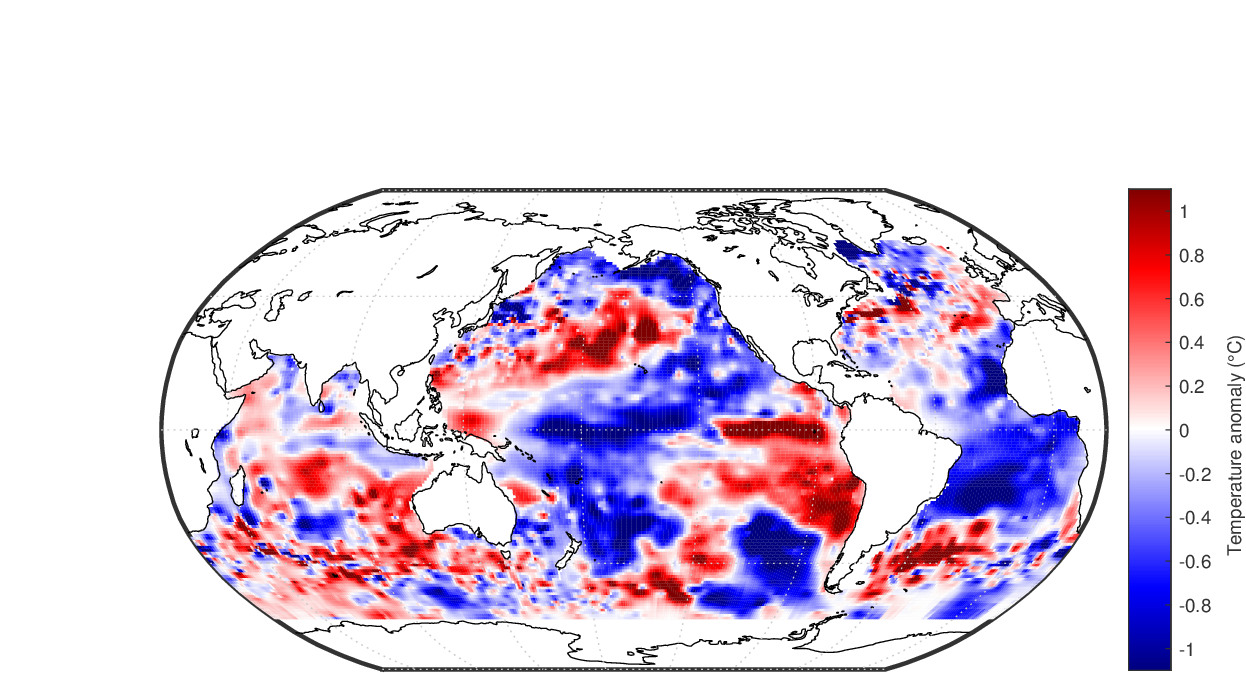}}
	\hfill
	\subfigure[10 dbar (reference)]{
		\includegraphics[width=7.7cm,trim = 5cm 0cm 0cm 6cm, clip=true]{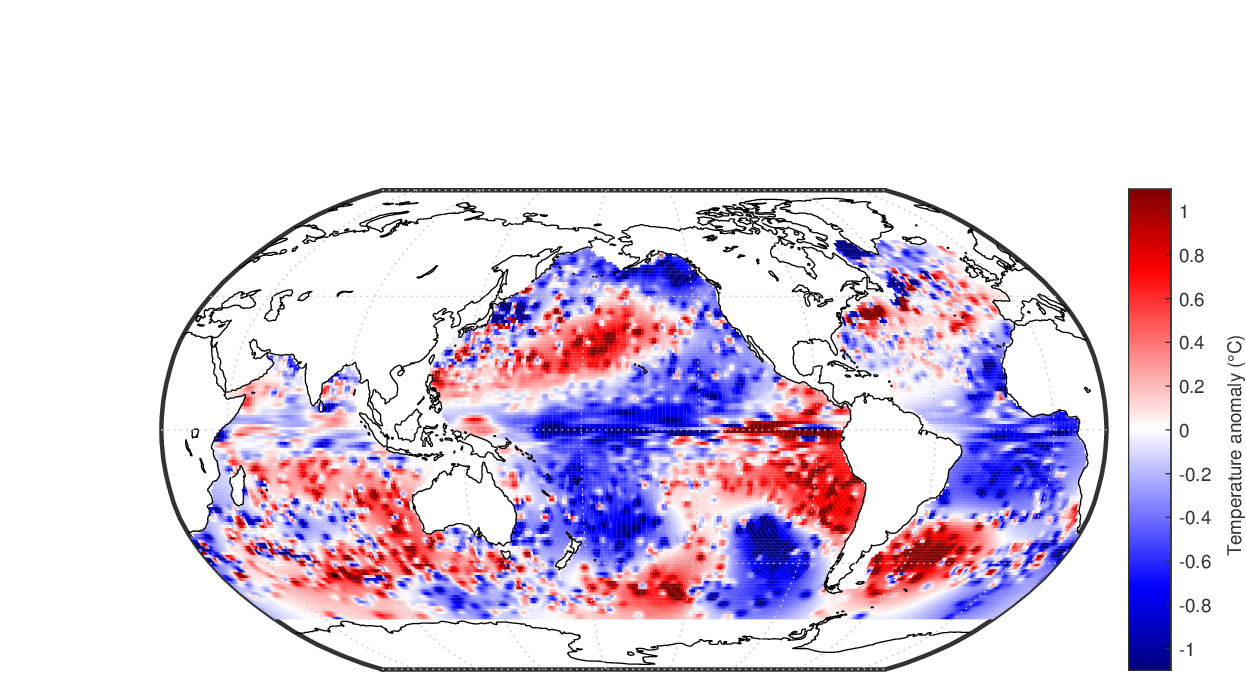}}
	\subfigure[300 dbar (local space-time)]{
		\includegraphics[width=7.7cm,trim = 5cm 0cm 0cm 6cm, clip=true]{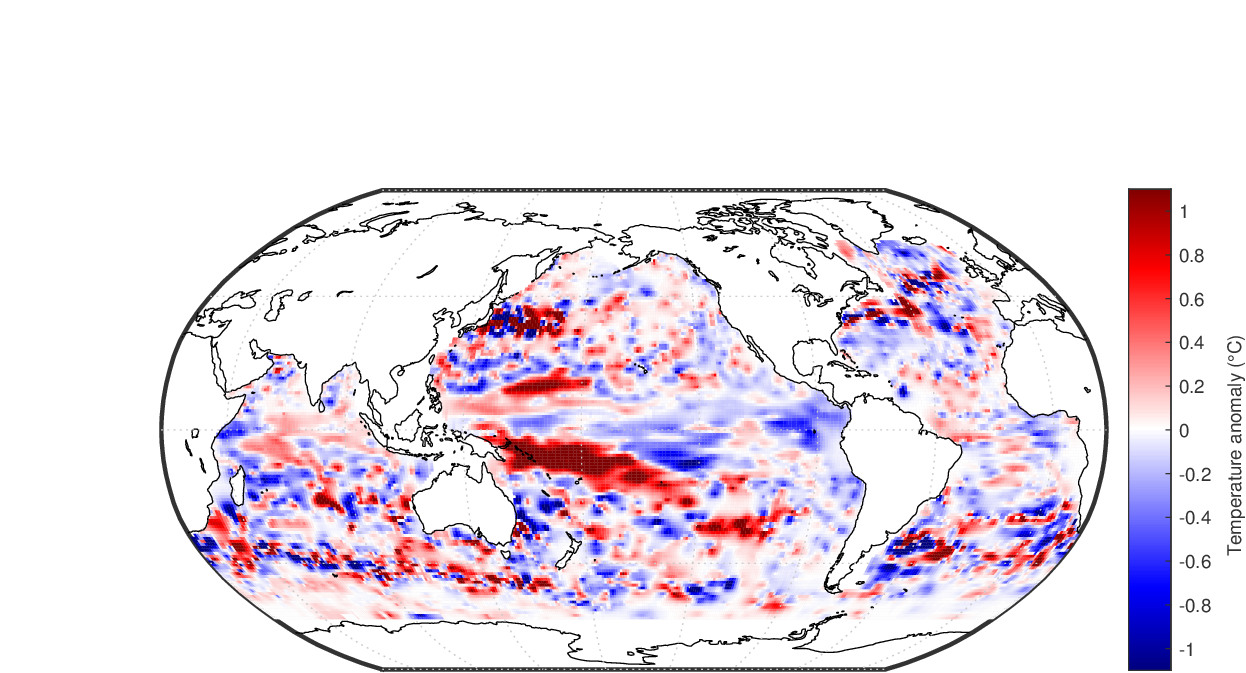}}
	\hfill
	\subfigure[300 dbar (reference)]{
		\includegraphics[width=7.7cm,trim = 5cm 0cm 0cm 6cm, clip=true]{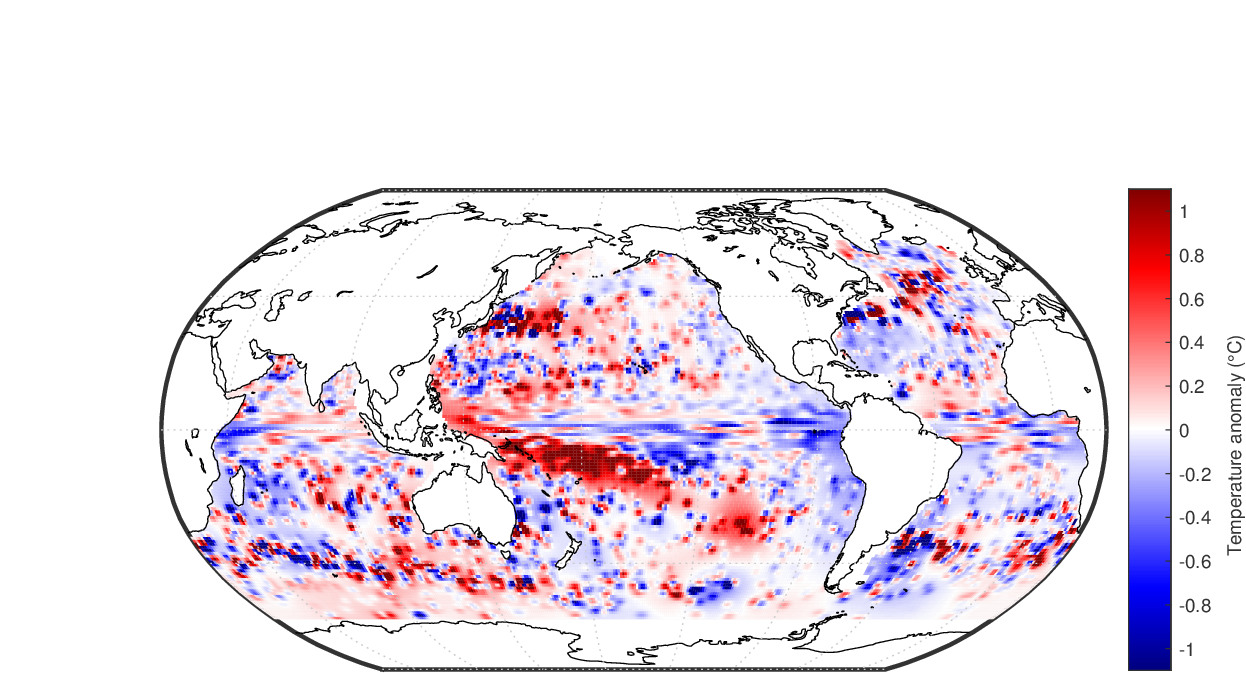}}
	\subfigure[1500 dbar (local space-time)]{
		\includegraphics[width=7.7cm,trim = 5cm 0cm 0cm 6cm, clip=true]{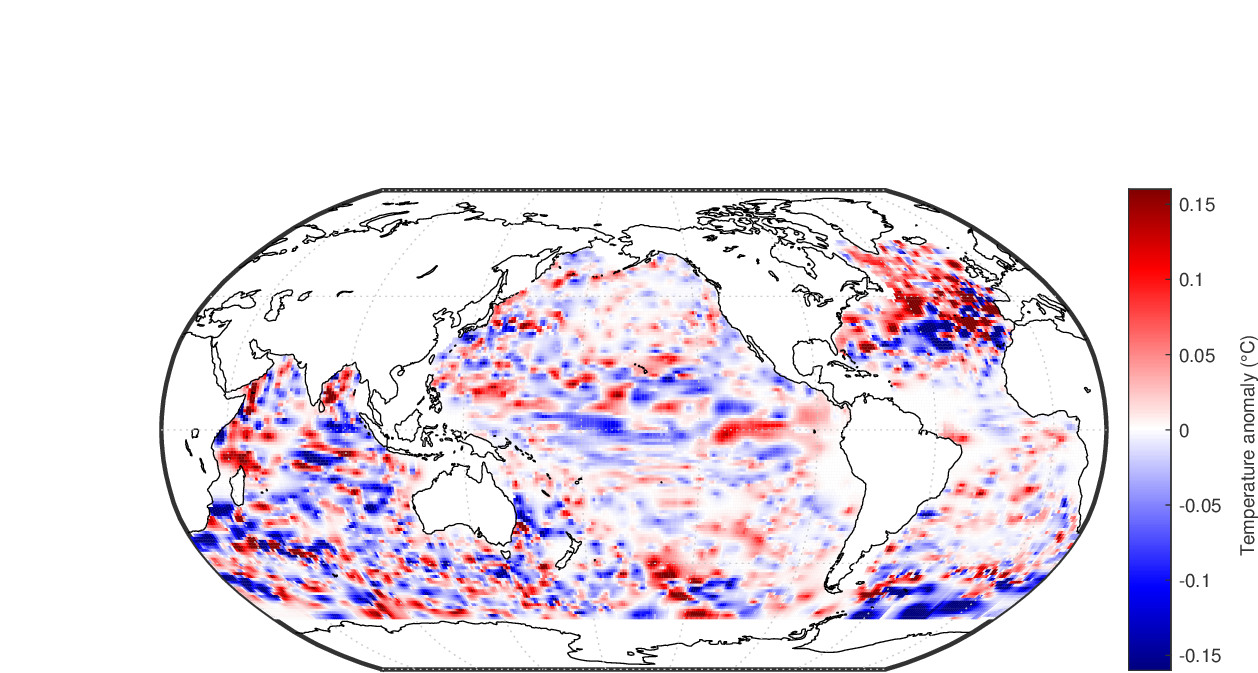}}
	\hfill
	\subfigure[1500 dbar (reference)]{
		\includegraphics[width=7.7cm,trim = 5cm 0cm 0cm 6cm, clip=true]{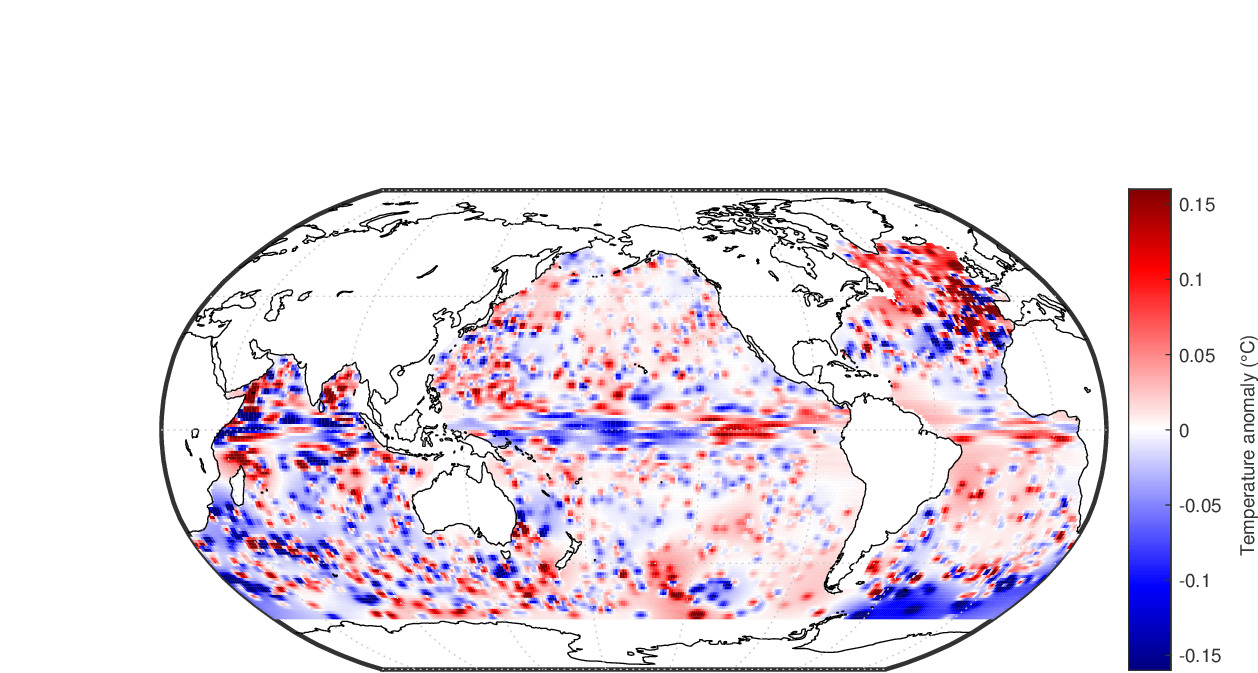}}
	
	\caption{Argo temperature anomalies for February 2012 at 10, 300 and 1500~dbar for the locally stationary 3-month spatio-temporal model with a Gaussian nugget (Model~5, left column) and for the Roemmich--Gilson-like reference model (Model~1, right column).}
	\label{fig:anomalies}
\end{figure}

We first investigate the performance of the different models in making point predictions of the temperature anomalies. Figure~\ref{fig:anomalies} displays the February 2012 temperature anomalies at 10, 300 and 1500~dbar for the locally stationary spatio-temporal model with a Gaussian nugget (Model~5) and for the reference model (Model 1). The maps are on a $1^\circ \times 1^\circ$ grid and the space-time field is evaluated at noon on February 15, 2012. The overall patterns in both fields are similar: one can recognize the large-scale anomalies near the surface, the elongated patterns in subsurface Equatorial regions and the meanders and eddies associated with the western boundary currents and the Antarctic Circumpolar Current. There are, however, a number of clear differences between the maps. Especially at 10~dbar, the reference map shows small speckles that are absent from the locally fitted map. This is related to the Roemmich--Gilson covariance parameters which are not optimized for mapping anomalies near the surface. Also, while present in both maps, the zonal elongation of the anomalies in the Equatorial regions is much more pronounced in the reference maps.

In order to study the difference between the reference method and the locally stationary maps in more quantitative terms, we perform a cross-validation study comparing the predictive performance of the different methods. We use two cross-validation strategies: in leave-one-observation-out (LOOO) cross-validation, we leave out one temperature observation at a time, while in leave-one-float-out (LOFO) cross-validation we leave out an entire float. LOOO predictions are easier to make since there will almost always be nearby observations from the same float to constrain the temperature value at the prediction location. LOFO cross-validation, on the other hand, creates a ``gap'' in the Argo array and has a higher amount of irreducible error. In the actual mapping problem, the typical distance from a grid point to nearby floats will be somewhere between these two extremes.

We investigate the cross-validation performance in terms of the root-mean-square error (RMSE), the median absolute error (MdAE) and the third quartile of the absolute error (Q$_3$AE). Let $\hat{y}_{-(i,j)}$ be the prediction at $(\bm{x}_{i,j},t_{i,j})$ with either $y_{i,j}$ removed from the dataset (LOOO) or all observations with the same float ID as $y_{i,j}$ removed (LOFO). Then
\begin{equation}
	\text{RMSE} = \sqrt{\frac{1}{\sum_{i=1}^n m_i} \sum_{i=1}^n \sum_{j=1}^{m_i} (\hat{y}_{-(i,j)}-y_{i,j})^2}
\end{equation}
and MdAE and Q$_3$AE are the sample median and sample third quartile of $\{|\hat{y}_{-(i,j)}-y_{i,j}|,\; i=1,\ldots,n,\; j=1,\ldots,m_i\}$. When cross-validating $y_{i,j}$, the model parameters are taken from the moving window centered at the $1^\circ \times 1^\circ$ grid point closest to $y_{i,j}$. The parameter estimates and the mean fields are kept fixed during the cross-validation.

Table~\ref{tab:CV_LOOO_Gaussian} summarizes the LOOO cross-validation performance for the models with a Gaussian nugget (Models~1, 2, 4 and 5). Also included is the performance when the predictions are made using only the Roemmich--Gilson spatial mean without any modeling of the anomalies (the performance of the spatio-temporal mean is only slightly better and is omitted for clarity). For all three pressure levels and for all three performance metrics, the relative performance of the models is in the following order: the reference model (Model~1) outperforms the spatial mean, the locally stationary 1-month spatial model (Model~2) outperforms the reference model (Model~1), the locally stationary 3-month spatial model (Model~4) outperforms the 1-month spatial model (Model~2) and the locally stationary 3-month spatio-temporal model (Model~5) outperforms the 3-month spatial model (Model~4). The combination of data-driven local covariance parameters, a larger temporal window and a covariance structure that includes time leads to a fairly substantial 10\:\% -- 30\:\% performance improvement over the reference model. The improvement is particularly large near the surface at 10~dbar. The distribution of the squared prediction errors (not shown) has a much fatter right tail than there would be if the errors followed a common normal distribution, so we also consider MdAE and Q$_3$AE as summaries of prediction performance to make sure that our conclusions are not driven by a small fraction of poor predictions. Table~\ref{tab:CV_LOOO_Gaussian} confirms that all three performance metrics are consistent in their relative ranking of the various methods.

\begin{table}[t]
	\caption{Point prediction performance measured in terms of the root-mean-square error (RMSE), the third quartile of the absolute error (Q$_3$AE) and the median absolute error (MdAE) for leave-one-observation-out (LOOO) cross-validation. The compared models are the Roemmich--Gilson spatial mean, the Roemmich--Gilson-like reference model (Model~1), the locally stationary 1-month and 3-month spatial models (Models~2 and 4) and the locally stationary 3-month spatio-temporal model (Model~5), all with a Gaussian nugget. The units are degrees Celsius and the percentages in the parentheses are improvements in comparison to the reference model.}
	\label{tab:CV_LOOO_Gaussian}
	\centering
	\begin{tabular}{lllllll}
		\toprule
		Pressure & Perf. & Spatial & Reference & Space & Space & Space-time \\
		level & metric & mean & model & (1 month) & (3 months) & (3 months) \\
		\midrule
		10 dbar & RMSE & 0.8889 & 0.6135 & 0.5876 (4.2\,\%) & 0.5667 (7.6\,\%) & 0.5072 (17.3\,\%) \\
		& Q$_3$AE & 0.8670 & 0.5026 & 0.4824 (4.0\,\%) & 0.4568 (9.1\,\%) & 0.3735 (25.7\,\%) \\
		& MdAE & 0.4750 & 0.2556 & 0.2490 (2.6\,\%) & 0.2293 (10.3\,\%) & 0.1801 (29.5\,\%) \\
		\midrule
		300 dbar & RMSE & 0.8149 & 0.5782 & 0.5692 (1.6\,\%) & 0.5675 (1.9\,\%) & 0.5124 (11.4\,\%) \\
		& Q$_3$AE & 0.6320 & 0.4213 & 0.4150 (1.5\,\%) & 0.4005 (4.9\,\%) & 0.3684 (12.6\,\%) \\
		& MdAE & 0.3062 & 0.1991 & 0.1957 (1.7\,\%) & 0.1873 (5.9\,\%) & 0.1740 (12.6\,\%) \\
		\midrule
		1500 dbar & RMSE & 0.1337 & 0.1014 & 0.0997 (1.7\,\%) & 0.0935 (7.8\,\%) & 0.0883 (12.9\,\%) \\
		& Q$_3$AE & 0.1043 & 0.0736 & 0.0725 (1.5\,\%) & 0.0678 (7.9\,\%) & 0.0641 (12.8\,\%) \\
		& MdAE & 0.0530 & 0.0356 & 0.0355 (0.3\,\%) & 0.0328 (8.0\,\%) & 0.0311 (12.7\,\%) \\
		\bottomrule 
	\end{tabular}
\end{table}

\begin{table}[t]
	\caption{Same as Table~\ref{tab:CV_LOOO_Gaussian} but for leave-one-float-out (LOFO) cross-validation.}
	\label{tab:CV_LOFO_Gaussian}
	\centering
	\begin{tabular}{lllllll}
		\toprule
		Pressure & Perf. & Spatial & Reference & Space & Space & Space-time \\
		level & metric & mean & model & (1 month) & (3 months) & (3 months) \\
		\midrule
		10 dbar & RMSE & 0.8889 & 0.7177 & 0.6823 (4.9\,\%) & 0.6954 (3.1\,\%) & 0.6489 (9.6\,\%) \\
		& Q$_3$AE & 0.8670 & 0.6107 & 0.5776 (5.4\,\%) & 0.5987 (2.0\,\%) & 0.5222 (14.5\,\%) \\
		& MdAE & 0.4750 & 0.3165 & 0.2981 (5.8\,\%) & 0.3062 (3.2\,\%) & 0.2552 (19.3\,\%) \\
		\midrule
		300 dbar & RMSE & 0.8149 & 0.7686 & 0.7486 (2.6\,\%) & 0.7483 (2.6\,\%) & 0.7388 (3.9\,\%) \\
		& Q$_3$AE & 0.6320 & 0.5942 & 0.5733 (3.5\,\%) & 0.5666 (4.6\,\%) & 0.5556 (6.5\,\%) \\
		& MdAE & 0.3062 & 0.2856 & 0.2753 (3.6\,\%) & 0.2732 (4.3\,\%) & 0.2664 (6.7\,\%) \\
		\midrule
		1500 dbar & RMSE & 0.1337 & 0.1373 & 0.1308 (4.8\,\%) & 0.1313 (4.4\,\%) & 0.1307 (4.8\,\%) \\
		& Q$_3$AE & 0.1043 & 0.1015 & 0.0976 (3.9\,\%) & 0.0973 (4.2\,\%) & 0.0959 (5.6\,\%) \\
		& MdAE & 0.0530 & 0.0511 & 0.0499 (2.3\,\%) & 0.0491 (3.7\,\%) & 0.0484 (5.3\,\%) \\
		\bottomrule 
	\end{tabular}
\end{table}

With LOFO cross-validation (Table~\ref{tab:CV_LOFO_Gaussian}), the prediction errors are consistently larger than with LOOO cross-validation, which reflects the more challenging nature of the LOFO prediction task. Even in this case, there are still distinct advantages from appropriate modeling of the anomalies. The ranking of the models and the general conclusions are otherwise the same as above, except for two differences: First, here the 3-month spatial model (Model~4) does not significantly improve upon the 1-month model (Model~2) and can in fact even perform worse. This happens because the model confuses spatial and temporal variation, which highlights the importance of using a full spatio-temporal covariance model. Second, at 1500~dbar, the RMSE of the reference model is slightly larger than the RMSE of the spatial mean. As discussed above, this may happen because of a few values in the right tail of the squared prediction error distribution, but may also indicate that the Roemmich--Gilson covariance parameters are not particularly well-suited for this pressure level. In comparison, all the data-driven models perform better than the spatial mean, as~expected.

The cross-validation results for the Student nugget are given in the supplement \cite{Kuusela2017Supplement}. The Student models tend to perform worse than the comparable Gaussian models. This happens because they smooth out non-Gaussian high frequency features (eddies in particular) by including them in the nugget term. Nevertheless, the same conclusion that the spatio-temporal model outperforms the purely spatial model remains true. Even though the Student models have inferior point prediction performance, they offer significant advantages in uncertainty quantification (see Section~\ref{sec:results}\ref{sec:UQ}).

To summarize, these results highlight the importance of including time in the mapping and allowing the covariance parameters to change with location and pressure. Indeed, the largest improvements are observed at 10~dbar, where one would intuitively expect the data-driven covariances to differ a lot from the Roemmich--Gilson model (see Section~\ref{sec:background}\ref{sec:RG_climatology}). To put these results into perspective, it is useful to keep in mind that statistical procedures typically converge at sublinear rates as the amount of data increases. For example, assuming a $1/\sqrt{n}$ rate of convergence, a $20\,\%$ improvement in the predictive performance translates into $56\,\%$ more data. This would correspond to deploying roughly 2,000 additional floats at a cost of 30 million~USD (one Argo float costs approximately 15,000~USD \cite{ArgoFAQ2017}). Similarly, even a $10\,\%$ performance improvement corresponds to approximately $23\,\%$ more data at a cost of some 12 million~USD. It should be noted that these dollar amounts are rough order-of-magnitude estimates using Argo-type floats. The actual costs would be higher if one took into account the need to replenish the array and the cost of data handling. On the other hand, improved upper ocean sampling could also be achieved at a lower cost by deploying cheaper floats that sample only the upper few hundred meters of the water column.

\subsection{Uncertainty quantification} \label{sec:UQ}

\begin{figure}[t]
	\centering
	
	\subfigure[10 dbar (local space-time)]{
		\includegraphics[width=7.6cm,trim = 5cm 0cm 0cm 6cm, clip=true]{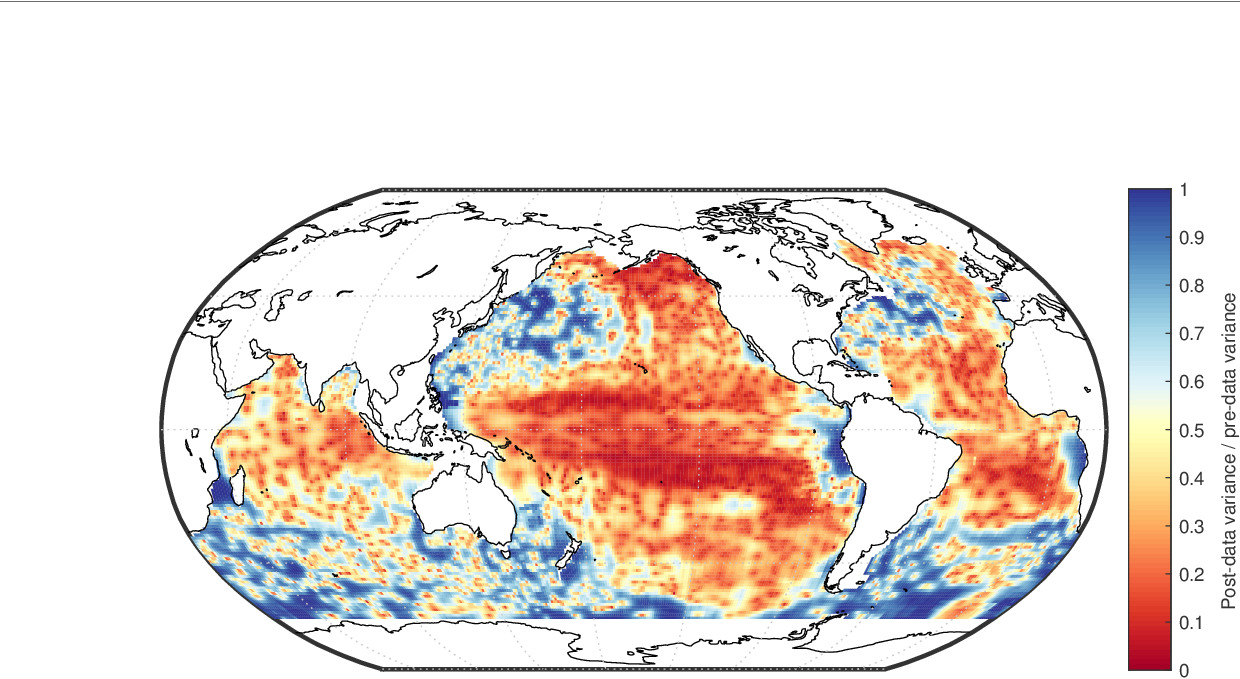} \label{fig:postDataVsPreDataSpaceTimeExp10}}
	\subfigure[300 dbar (local space-time)]{
		\includegraphics[width=7.6cm,trim = 5cm 0cm 0cm 6cm, clip=true]{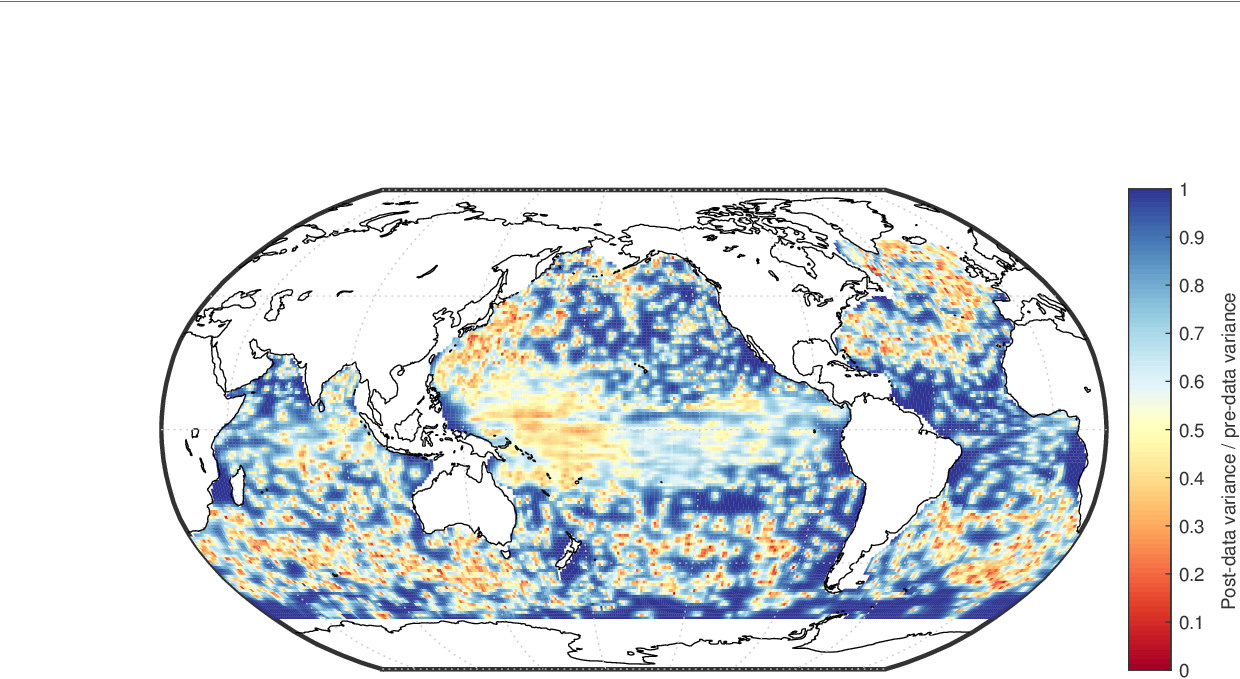} \label{fig:postDataVsPreDataSpaceTimeExp300}}
	\subfigure[1500 dbar (local space-time)]{
		\includegraphics[width=7.6cm,trim = 5cm 0cm 0cm 6cm, clip=true]{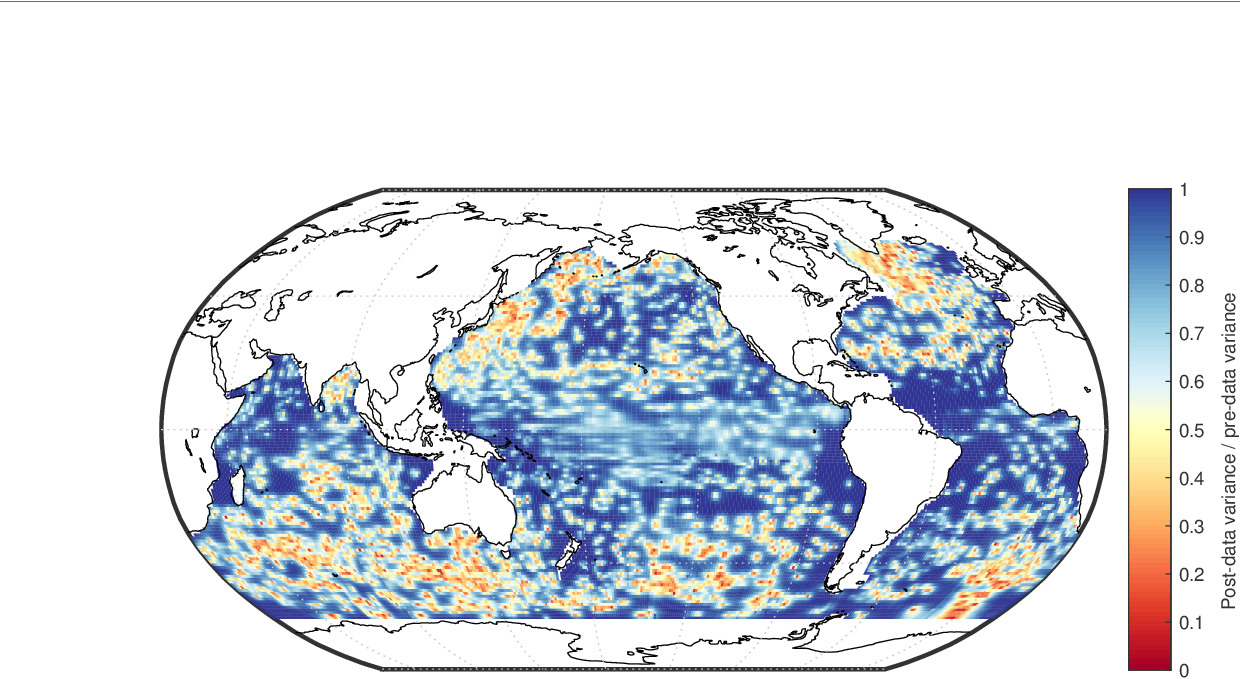} \label{fig:postDataVsPreDataSpaceTimeExp1500}}
	\subfigure[10 dbar (reference)]{
		\includegraphics[width=7.6cm,trim = 5cm 0cm 0cm 6cm, clip=true]{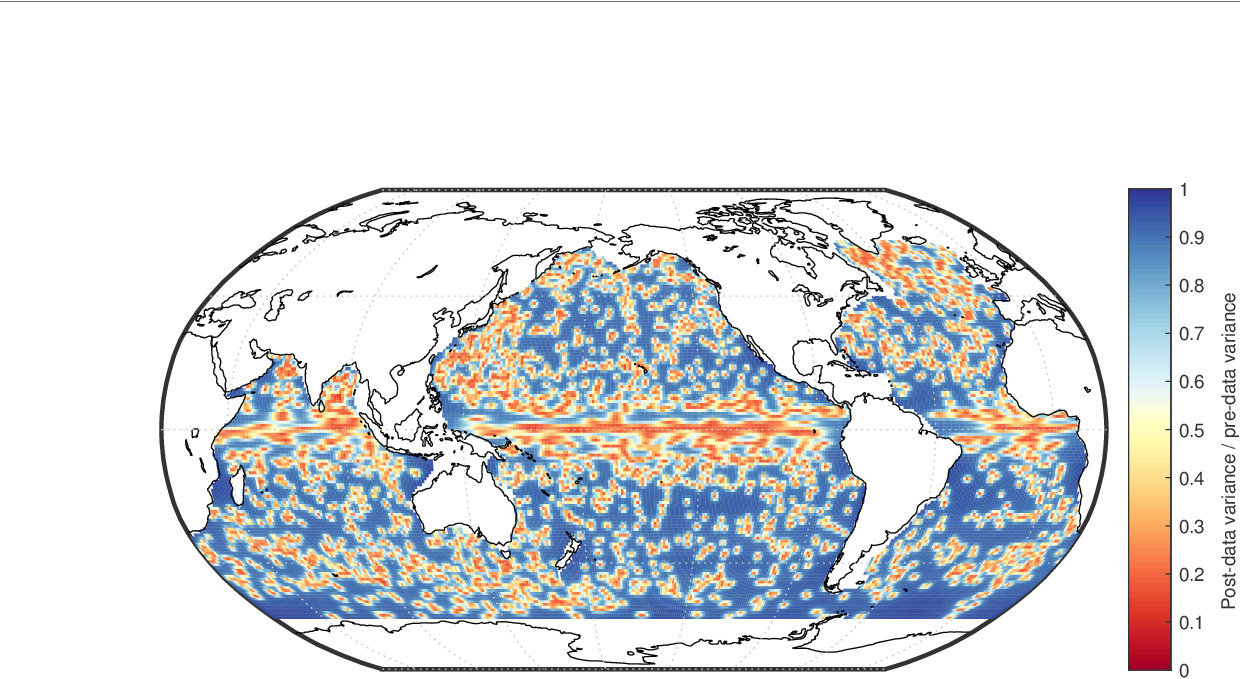} \label{fig:postDataVsPreDataRG}}
	
	\caption{Post-data-to-pre-data variance ratio $\var(y_i^*|\bm{y}_i,\bm{\hat{\theta}},\hat{\sigma}^2) \big/ \var(y_i^*|\bm{\hat{\theta}},\hat{\sigma}^2)$ in February 2012 for the locally stationary 3-month spatio-temporal model with a Gaussian nugget (Model~5) at 10, 300 and 1500~dbar (Figures~\subref{fig:postDataVsPreDataSpaceTimeExp10}\hbox{--}\subref{fig:postDataVsPreDataSpaceTimeExp1500}) and for the Roemmich--Gilson-like reference model (Model~1) at 10~dbar (Figure~\subref{fig:postDataVsPreDataRG}). Because the Roemmich--Gilson covariance is the same at all pressures, the variance ratios for the reference model at 300~dbar and 1500~dbar (not shown) are essentially the same as the one at 10~dbar.}
	\label{fig:postDataVsPreData}
\end{figure}

We next investigate the uncertainty quantification performance of the different models. Figure~\ref{fig:postDataVsPreData} displays the post-data-to-pre-data variance ratio $\var(y_i^*|\bm{y}_i,\bm{\hat{\theta}},\hat{\sigma}^2) \big/ \var(y_i^*|\bm{\hat{\theta}},\hat{\sigma}^2) = \big(\hat{\phi} + \hat{\sigma}^2 - \tp{(\bm{k}_i^*(\bm{\hat{\theta}}))}(\bm{K}_i(\bm{\hat{\theta}})+\hat{\sigma}^2\bm{I})^{-1} \bm{k}_i^*(\bm{\hat{\theta}})\big)\big/\big(\hat{\phi} + \hat{\sigma}^2\big)$, i.e., the ratio of the predictive variances with and without Argo data, for the locally stationary 3-month spatio-temporal model with a Gaussian nugget (Model~5) at 10, 300 and 1500~dbar in February 2012. When this ratio is close to 0, Argo data provide firm inferences about the temperature anomaly, while a value close to 1 means that observing Argo data did not considerably reduce our uncertainty about the temperature anomaly at that particular location and time. Also shown is the same ratio for the reference model (Model~1) at 10~dbar. Since the uncertainties of Gaussian process models depend only on the covariance parameters and the observation locations and since the Roemmich--Gilson covariance is the same at all pressure levels, the uncertainty of the reference model at 300~dbar and 1500~dbar (not shown) looks essentially the same as the uncertainty at 10~dbar (there are minor differences due to some profiles not extending all the way from 10~dbar to 1500~dbar). By contrast, the locally stationary uncertainties are vastly different at different pressures. This happens because the estimated covariance parameters vary significantly as a function of pressure (see Section \ref{sec:results}\ref{sec:localDependence}). Based on the anomalies shown in Figure~\ref{fig:anomalies}, it makes intuitive sense that the uncertainties near the surface should be quite different from the uncertainties at greater depths. Note also that in the Roemmich--Gilson model with its fixed noise-to-signal variance ratio $\sigma^2 / \phi = 0.15$, the post-data-to-pre-data variance ratio in Figure~\ref{fig:postDataVsPreDataRG} does not depend on how the GP variance $\phi$ is chosen. By contrast, the rest of the results in this section require an estimate of $\phi$ (see Section~\ref{sec:results}\ref{sec:setupAndData}).

\begin{figure}[t]
	\centering
	
	\includegraphics[width=9.4cm,trim = 0cm 0cm 0cm 0cm, clip=true]{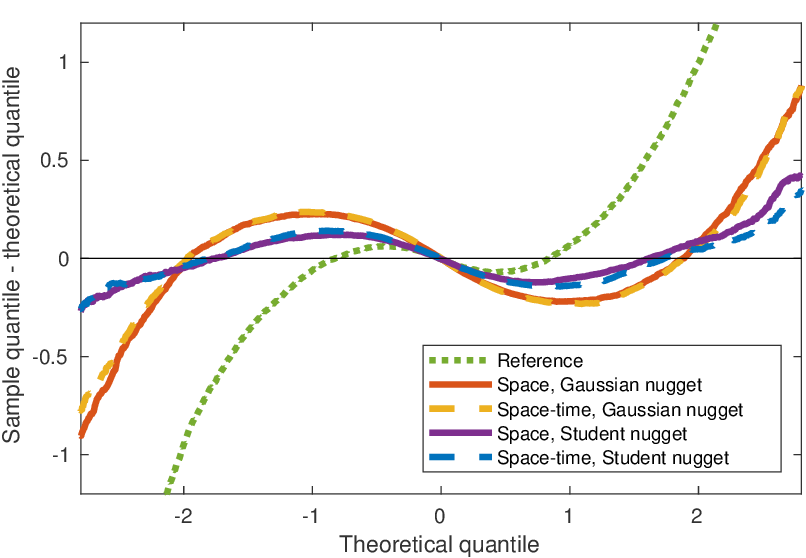}
	
	\caption{The difference of the cross-validated sample quantile and the corresponding standard Gaussian theoretical quantile ($q_\mathrm{sample} - q_\mathrm{theory}$) plotted against the theoretical quantile ($q_\mathrm{theory}$) for leave-one-observation-out (LOOO) cross-validation at 300~dbar. The compared models are the Roemmich--Gilson-like reference model (Model~1), the locally stationary 1-month spatial and 3-month spatio-temporal models with a Gaussian nugget (Models~2 and 5) and the corresponding models with a Student nugget (Models~3~and~6). The closer the curves are to a horizontal straight line at 0, the better the calibration of the predictive distributions.}
	\label{fig:qqPlot}
\end{figure}

In order to study the uncertainties in a more quantitative way, we cross-validate the entire predictive distribution for the different interpolation methods. We compare the calibration of the reference model (Model~1), the locally stationary 1-month spatial models with a Gaussian and a Student nugget (Models~2 and 3) and the locally stationary 3-month spatio-temporal models with a Gaussian and a Student nugget (Models~5 and 6). For each model, let $q_\mathrm{sample}$ denote the cross-validated predictive sample quantile on the $N(0,1)$ scale and $q_\mathrm{theory}$ the corresponding theoretical $N(0,1)$ quantile. The computation of $q_\mathrm{sample}$ is described in detail in the supplement~\cite{Kuusela2017Supplement}. We plot in Figure~\ref{fig:qqPlot} the quantile difference $q_\mathrm{sample}-q_\mathrm{theory}$ against the theoretical quantile $q_\mathrm{theory}$ at 300~dbar for LOOO cross-validation. This can be understood as the usual QQ plot with the identity line subtracted---plotting the quantiles this way helps visualize differences at the core of the distribution. The reference model is poorly calibrated with both tails of the data distribution much wider than those of the predictive distribution. The locally stationary models with the Gaussian nugget improve the calibration, but the data distribution still has heavier tails and a pointier core than the predictive distribution. We understand this as evidence of non-Gaussian heavy tails in the subsurface temperature data (similar heavy tails have been previously reported for sea surface temperatures; see \cite{Sura2008}). The Student nugget provides a way to account for these heavy tails and indeed the calibration of the Student models is much better than that of the fully Gaussian models. Even though some miscalibration still remains, the overall improvement over the reference model is quite substantial. We also note that the spatial and spatio-temporal models are essentially equally well-calibrated in the present~setting.

The supplement~\cite{Kuusela2017Supplement} stratifies Figure~\ref{fig:qqPlot} by years and latitude bins. This shows that February 2016 has slightly worse calibration than the other years, but otherwise the calibration is remarkably similar across the years, supporting the assumption made in Section~\ref{sec:methods} that the data can be treated as having the same distribution for each year. Stratification by latitude shows that the better calibration of the Student models primarily comes from improved performance in areas south of 45$^\circ$S and north of 45$^\circ$N.

The quantile plots in Figure~\ref{fig:qqPlot} translate directly into coverage probabilities for the predictive intervals. This is illustrated in Table~\ref{tab:coverage_LOOO_300dbar} which shows the LOOO cross-validated empirical coverages and interval lengths for 68\:\%, 95\:\% and 99\:\% predictive intervals at 300 dbar for the same models as in Figure~\ref{fig:qqPlot}. As expected based on Figure~\ref{fig:qqPlot}, the reference model undercovers at all three confidence levels. The locally stationary models overcover at 68\:\% level, are well-calibrated at 95\:\% level and undercover at 99\:\% level. At 68\:\% and 99\:\% levels, the coverage of the Student intervals is much closer to the nominal level than the coverage of the Gaussian intervals. The space-time intervals are always shorter than the corresponding spatial intervals, which is consistent with the performance improvements described in Section~\ref{sec:results}\ref{sec:pointPred}.

\begin{table}[!t]
	\caption{Empirical coverage and length (in ${}^\circ$C) of the predictive intervals at 300~dbar for leave-one-observation-out (LOOO) cross-validation. The models are the same as in Figure~\ref{fig:qqPlot}.}
	\label{tab:coverage_LOOO_300dbar}
	\centering
	\begin{tabular}{llllll}
		\toprule
		Confidence & Method & Empirical & Mean & Median \\
		level & & coverage & length & length \\
		\midrule
		68\:\% & Reference & 0.6607 & 0.7511 & 0.6435 \\
		& Space, Gaussian nugget & 0.7745 & 0.9749 & 0.8728 \\
		& Space-time, Gaussian nugget & 0.7800 & 0.8722 & 0.7816 \\
		& Space, Student nugget & 0.7261 & 0.8427 & 0.7621 \\
		& Space-time, Student nugget & 0.7389 & 0.7697 & 0.7049 \\
		\midrule
		95\:\% & Reference & 0.8755 & 1.4721 & 1.2612 \\
		& Space, Gaussian nugget & 0.9482 & 1.9108 & 1.7107 \\
		& Space-time, Gaussian nugget & 0.9490 & 1.7095 & 1.5320 \\
		& Space, Student nugget & 0.9432 & 1.9918 & 1.7525 \\
		& Space-time, Student nugget & 0.9452 & 1.8543 & 1.6192 \\
		\midrule
		99\:\% & Reference & 0.9329 & 1.9347 & 1.6575 \\
		& Space, Gaussian nugget & 0.9777 & 2.5112 & 2.2483 \\
		& Space-time, Gaussian nugget & 0.9793 & 2.2466 & 2.0134 \\
		& Space, Student nugget & 0.9835 & 3.9385 & 2.6417 \\
		& Space-time, Student nugget & 0.9844 & 3.8086 & 2.3976 \\
		\bottomrule 
	\end{tabular}
\end{table}

The quantile plots, empirical coverages and interval lengths for the other pressure levels and for LOFO cross-validation are given in the supplement \cite{Kuusela2017Supplement}. The LOOO conclusions at 10~dbar are similar to 300~dbar, except that the spatio-temporal models appear to have worse calibration than the spatial models. This might be an indication that the spatio-temporal dependence structure near the surface is more complicated than what our exponential covariance function with geometric anisotropy can capture. For LOOO at 1500~dbar, the spatial and spatio-temporal models are equally well-calibrated, but there also appears to be less non-Gaussianity and the Student nugget provides only limited improvement over the Gaussian nugget. The basic conclusions that locally stationary modeling improves the calibration over the reference model and that the Student nugget improves over the Gaussian nugget still largely hold true for LOFO cross-validation, but here the spatio-temporal models are slightly worse calibrated than the spatial models at all pressures. The spatio-temporal models nonetheless provide distinct advantages in terms of interval length. In these cases, the optimal choice of intervals depends on the relative importance given to accurate calibration and short interval length.

\subsection{Local estimates of dependence structure} \label{sec:localDependence}

In this section, we study the locally estimated model parameters and demonstrate that they exhibit physically meaningful patterns. We analyze in detail the 3-month spatio-temporal model with a Gaussian nugget (Model~5). Further plots and analogous results for the other models are given in the supplement~\cite{Kuusela2017Supplement}.

\begin{figure}[t]
	\centering
	
	\subfigure[300 dbar]{
		\includegraphics[width=8.5cm,trim = 5cm 0cm 0cm 6cm, clip=true]{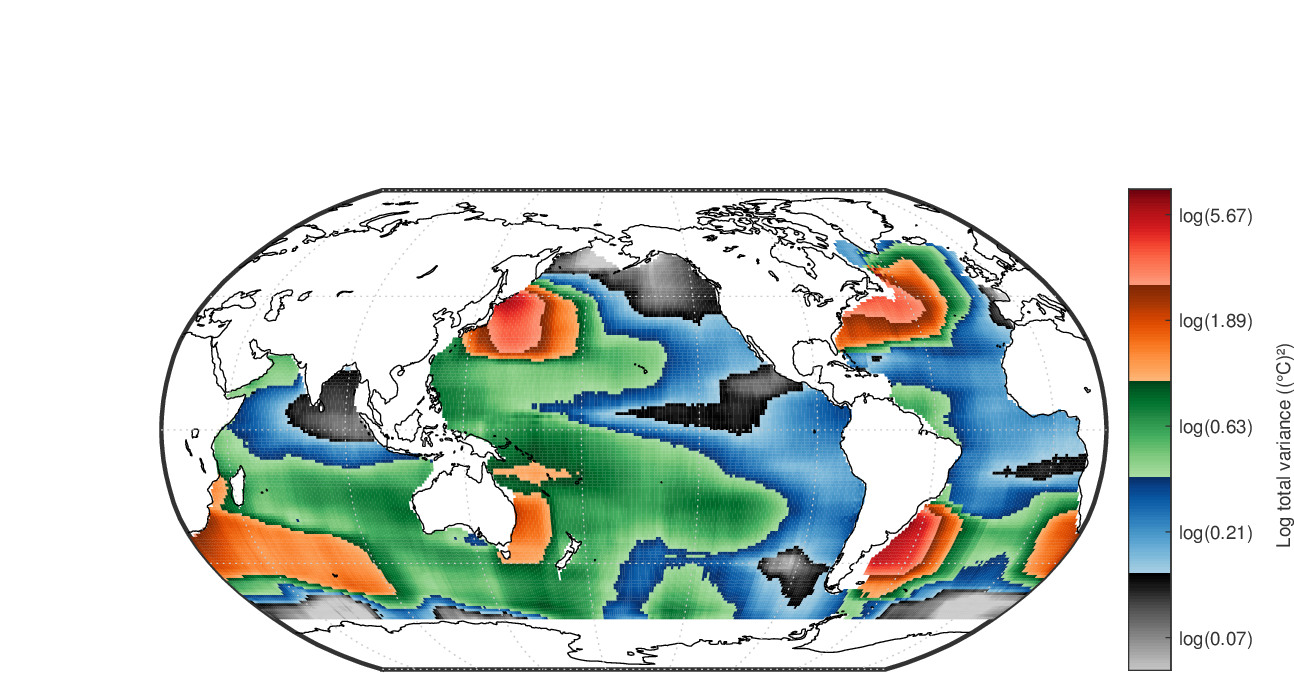}}\\
	\subfigure[10 dbar]{
		\includegraphics[width=7.7cm,trim = 5cm 0cm 0cm 6cm, clip=true]{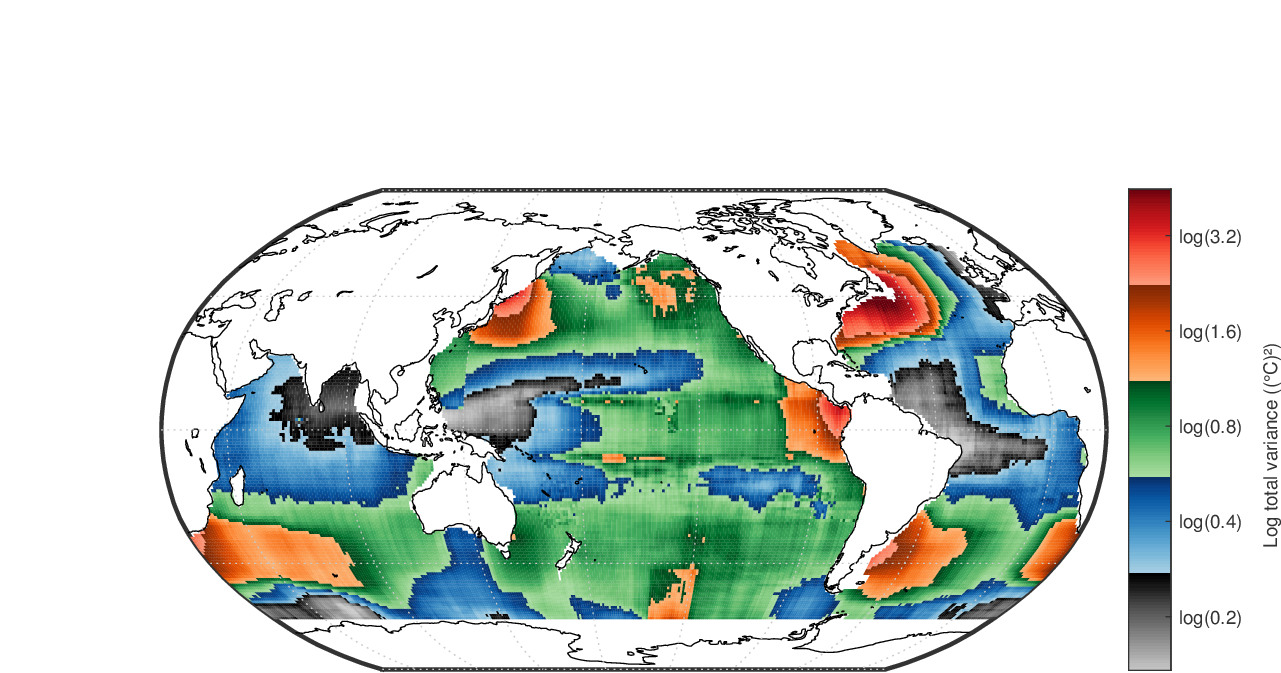}}
	\subfigure[1500 dbar]{
		\includegraphics[width=7.7cm,trim = 5cm 0cm 0cm 6cm, clip=true]{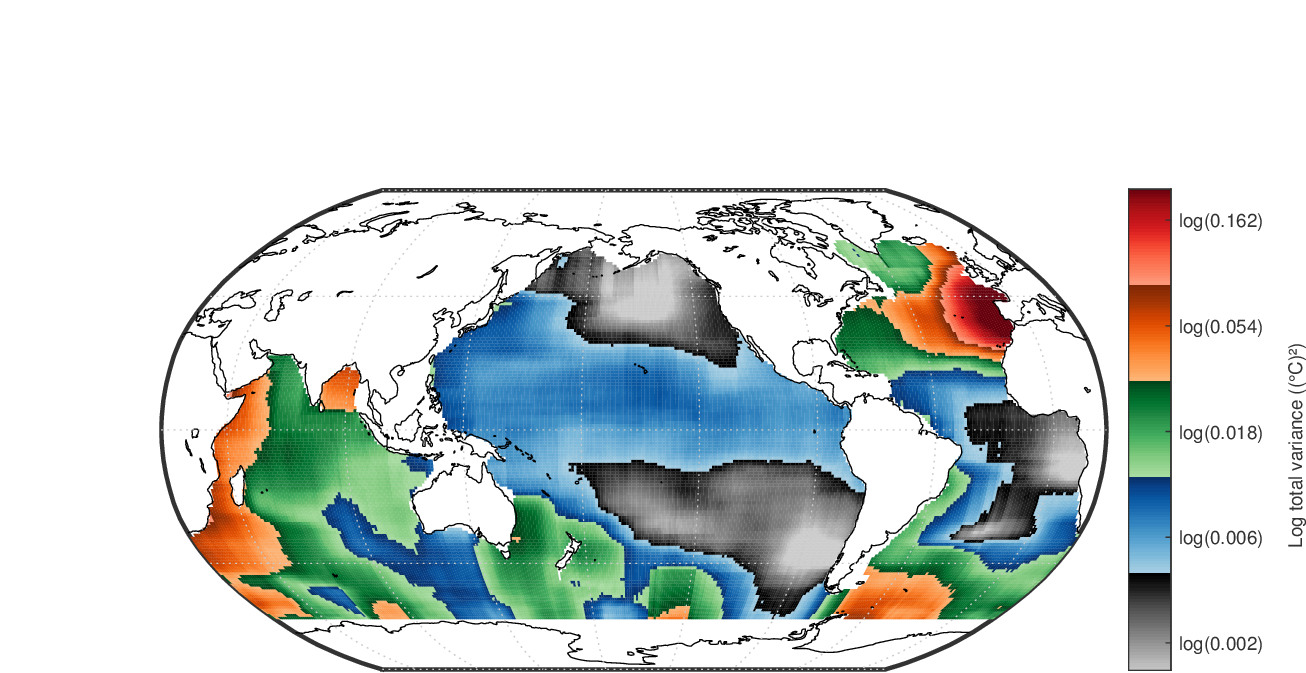}}
	
	\caption{Estimated total variance $\hat{\phi} + \hat{\sigma}^2$ for the locally stationary 3-month spatio-temporal model with a Gaussian nugget (Model~5).}
	\label{fig:thetasPlusSigma2}
\end{figure}

We first investigate the estimated total variance $\hat{\phi} + \hat{\sigma}^2$ at 10, 300 and 1500~dbar (Figure~\ref{fig:thetasPlusSigma2}). At the first two pressure levels, we can clearly see the large variability in western sides of the ocean basins caused by strong western boundary currents. For example, the Kuroshio Current off the coast of Japan, the Gulf Stream in the northwest Atlantic, the Brazil Current in the southwest Atlantic and the Agulhas retroflection and leakage areas around the southern tip of Africa (see Section~11.4.2 in~\cite{Talley2011}) can all be easily identified at both pressure levels. The East Australian Current is also visible at 300~dbar. By contrast, eastern sides of the ocean basins have significantly less variability, as expected. A notable exception is the northeastern Atlantic at 1500~dbar which has much more variability than other regions at this pressure level. We suspect that this can be attributed to eddies caused by the outflow from the Mediterranean Sea \cite{Iorga1999}. The bands of large variability at roughly 10$^\circ$N and 10$^\circ$S at 10~dbar in the Pacific Ocean might be related to Rossby waves at those~latitudes.

We next study the estimated ranges of zonal, meridional and temporal dependence in Argo temperature data. We do this by plotting maps of the correlation implied by the locally estimated covariance parameters at given zonal, meridional and temporal lags. We prefer to plot the maps in the correlation space instead of the range parameter space since there is some degree of ambiguity with regard to what fraction of the variability is attributed to the nugget effect in the MLE fits (at intermediate lags, a large nugget variance $\sigma^2$, a small GP variance $\phi$ and a large range parameter $\theta_\mathrm{lat}$, $\theta_\mathrm{lon}$ or $\theta_t$ can imply almost the same fitted correlation as a small nugget variance $\sigma^2$, a large GP variance $\phi$ and a small range parameter $\theta_\mathrm{lat}$, $\theta_\mathrm{lon}$ or $\theta_t$). The interested reader can find maps of the individual model parameters in the supplement \cite{Kuusela2017Supplement}.

\begin{figure}[t]
	\centering
	
	\subfigure[10 dbar, $\Delta x_\mathrm{lon} = 800$~km]{
		\includegraphics[width=7.7cm,trim = 5cm 0cm 0cm 6cm, clip=true]{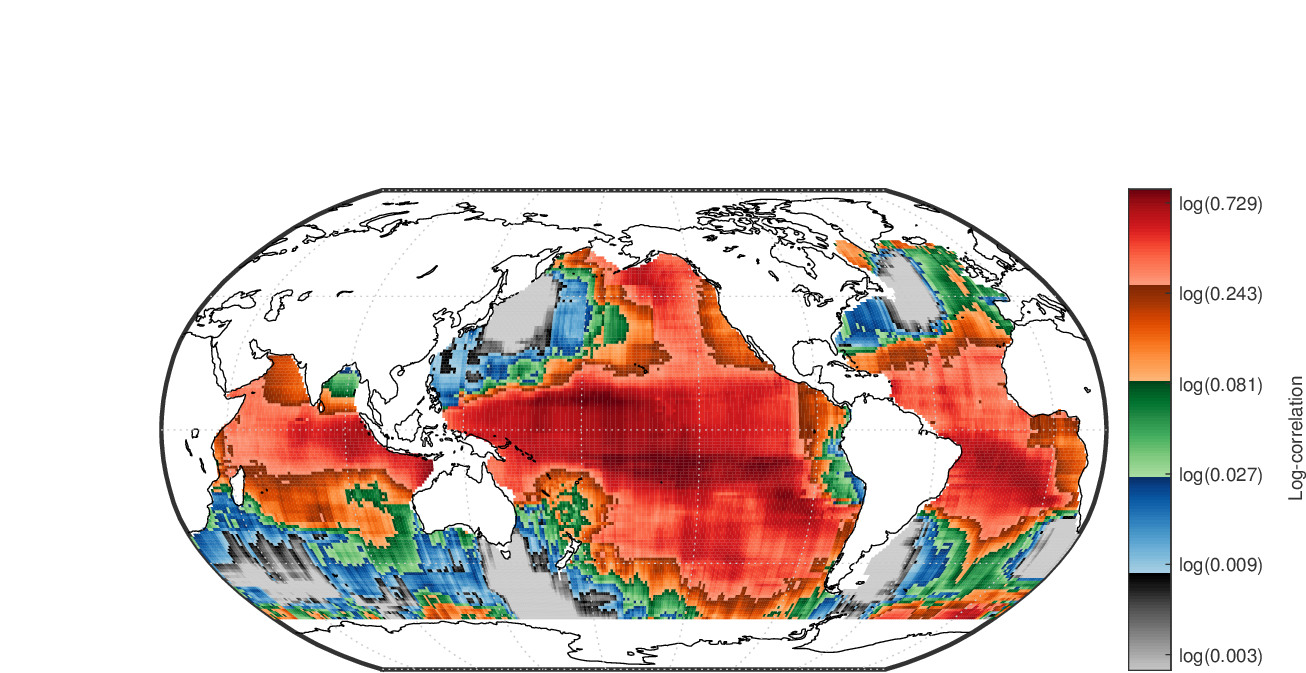}}
	\hfill
	\subfigure[10 dbar, $\Delta x_\mathrm{lat} = 800$~km]{
		\includegraphics[width=7.7cm,trim = 5cm 0cm 0cm 6cm, clip=true]{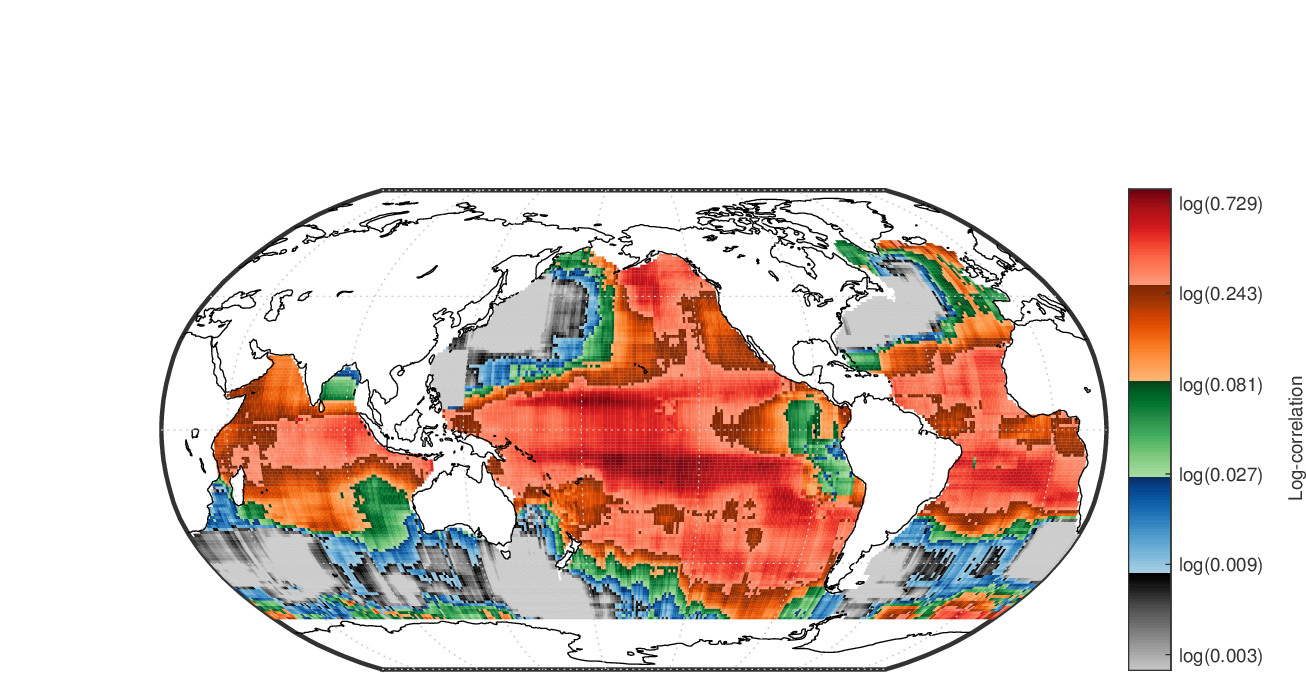}}
	\subfigure[300 dbar, $\Delta x_\mathrm{lon} = 800$~km]{
		\includegraphics[width=7.7cm,trim = 5cm 0cm 0cm 6cm, clip=true]{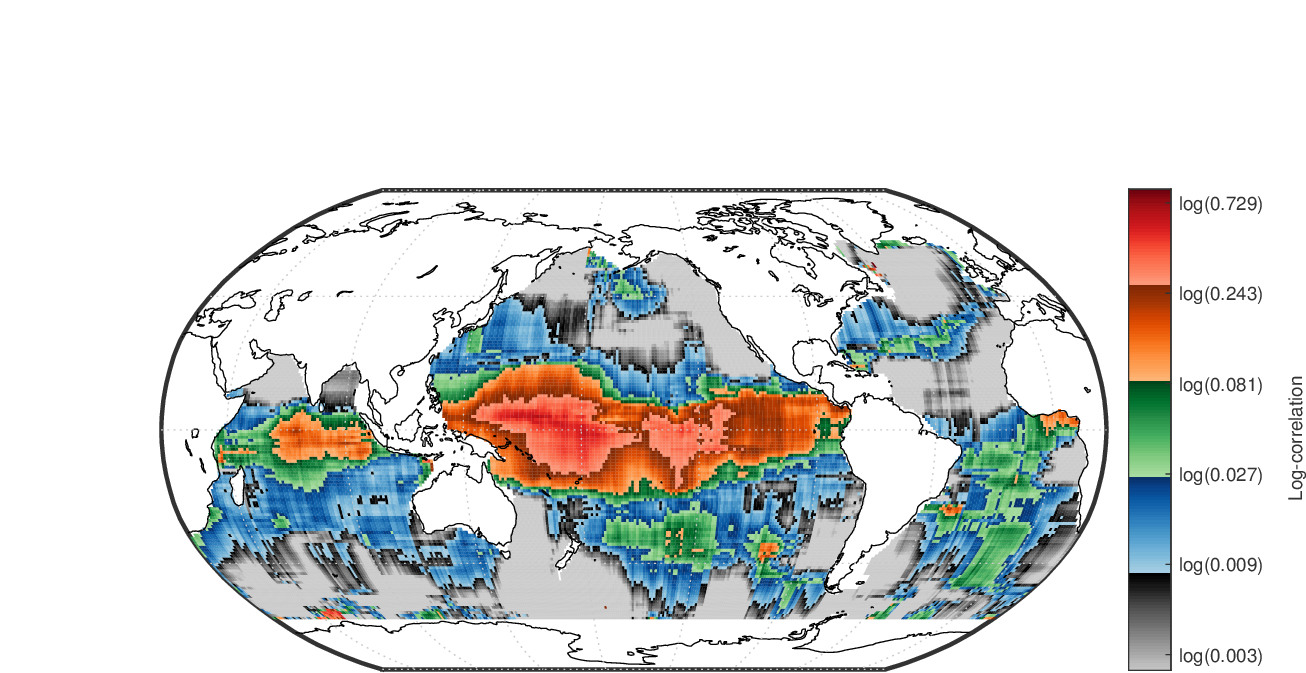}}
	\hfill
	\subfigure[300 dbar, $\Delta x_\mathrm{lat} = 800$~km]{
		\includegraphics[width=7.7cm,trim = 5cm 0cm 0cm 6cm, clip=true]{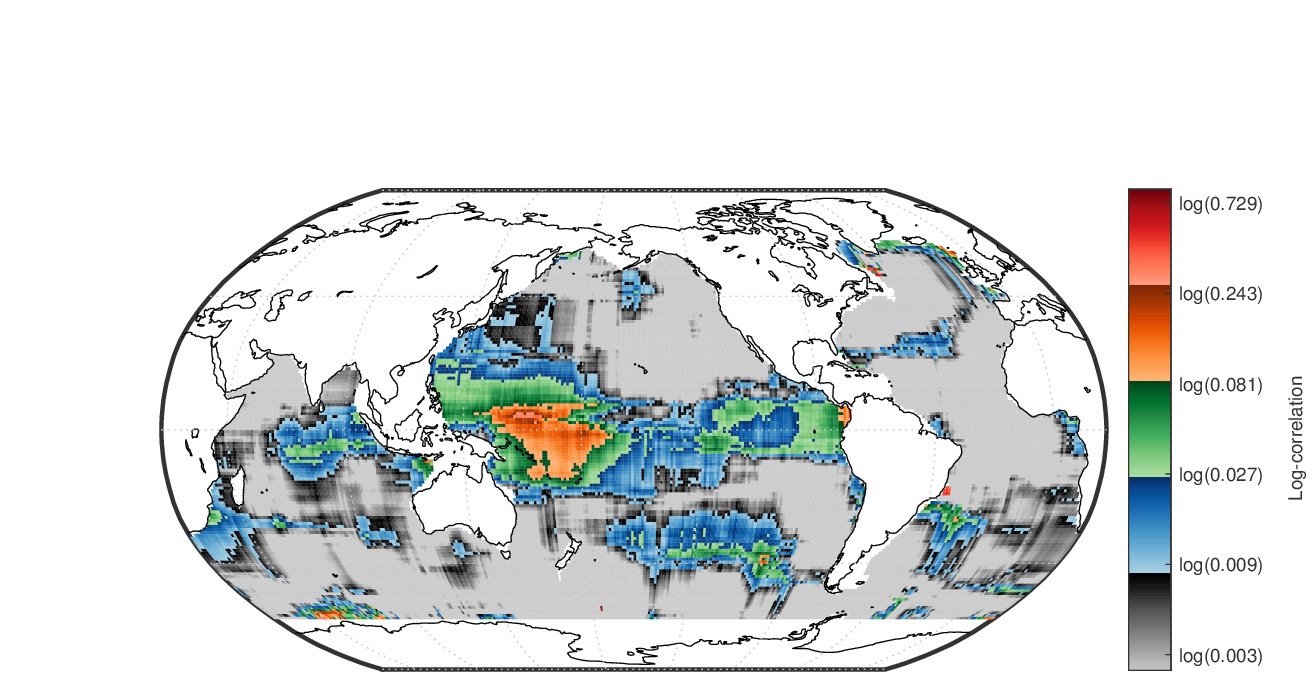}}
	\subfigure[1500 dbar, $\Delta x_\mathrm{lon} = 800$~km]{
		\includegraphics[width=7.7cm,trim = 5cm 0cm 0cm 6cm, clip=true]{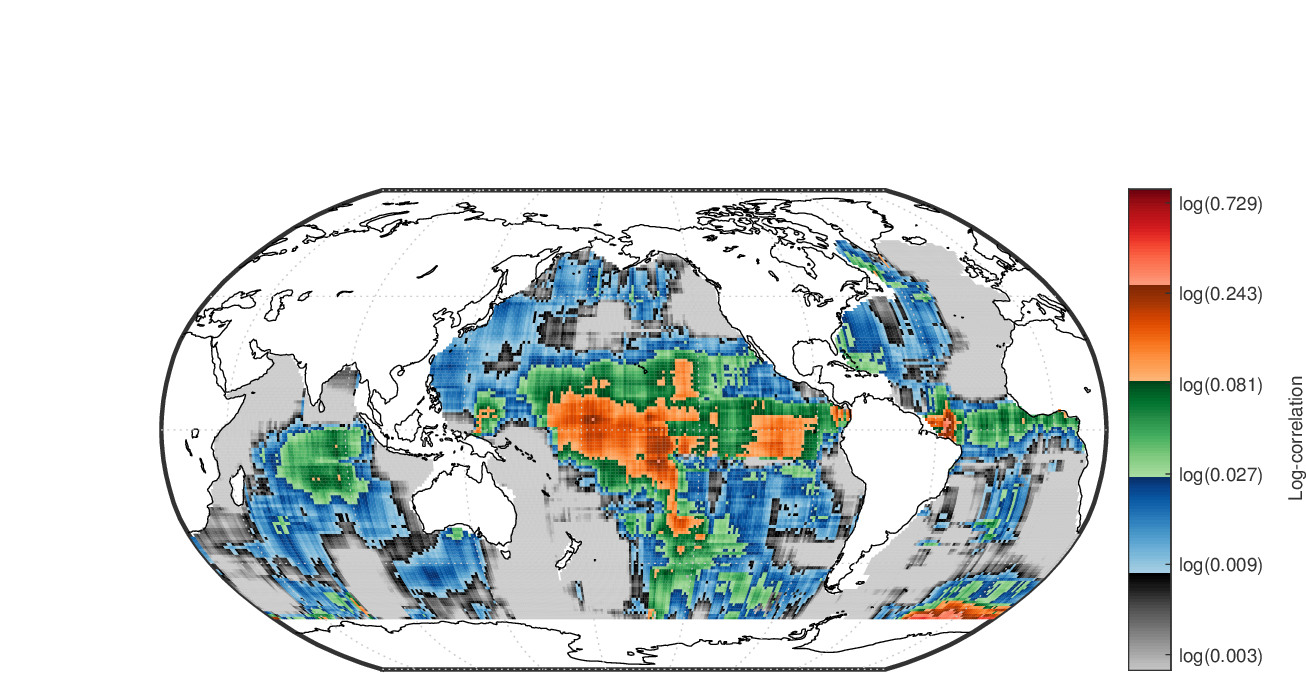}}
	\hfill
	\subfigure[1500 dbar, $\Delta x_\mathrm{lat} = 800$~km]{
		\includegraphics[width=7.7cm,trim = 5cm 0cm 0cm 6cm, clip=true]{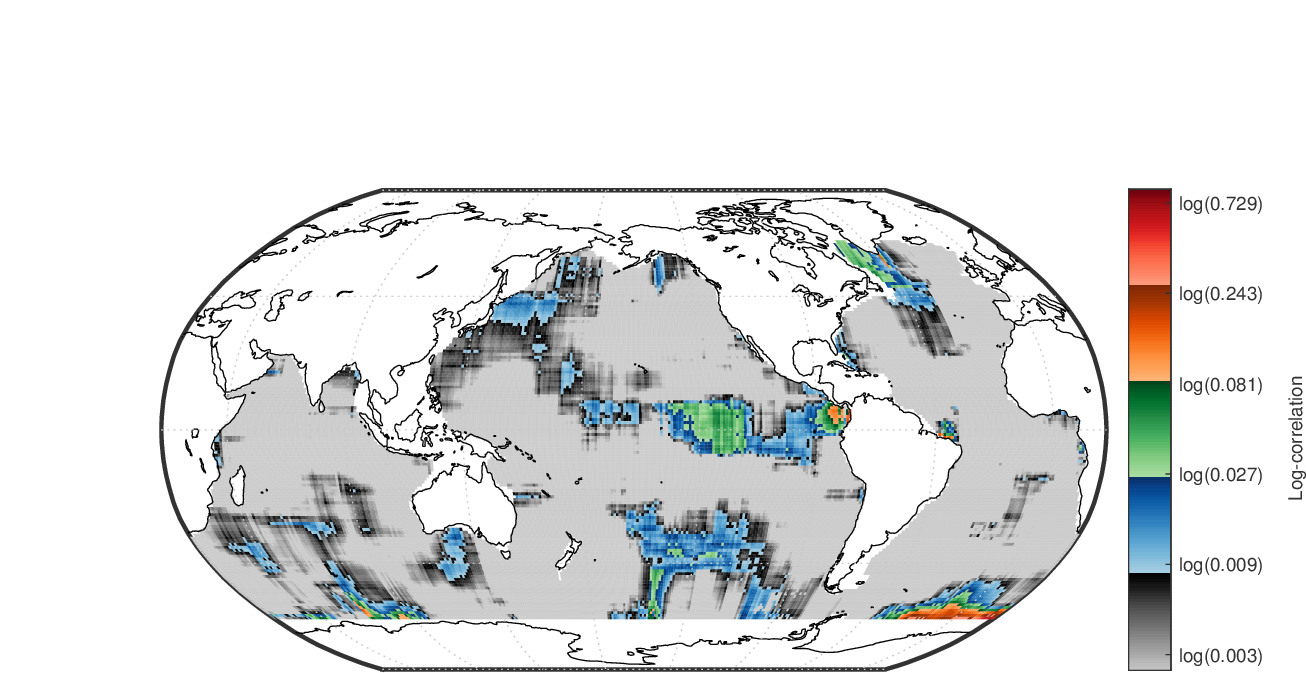}}
	
	\caption{Fitted correlations at zonal lag $\Delta x_\mathrm{lon} = 800$~km (left panels) and at meridional lag $\Delta x_\mathrm{lat} = 800$~km (right panels) for the locally stationary 3-month spatio-temporal model with a Gaussian nugget (Model~5). To facilitate comparison, all the panels have the same color scale.}
	\label{fig:corrLonLat}
\end{figure}

The left panels in Figure~\ref{fig:corrLonLat} show the fitted correlations at the zonal lag $\Delta x_\mathrm{lon} = 800$~km (with $\Delta x_\mathrm{lat} = 0$, $\Delta t = 0$ and taking the conversion latitude into account when converting degrees into kilometers). The range of zonal dependence varies considerably as a function of pressure, with much longer ranges near the surface than at greater depths. The large correlation values near the surface are most likely caused by interaction with the atmosphere. There is also a general tendency to have zonally elongated ranges in the Equatorial regions and the estimated ranges are generally longer in the Pacific Ocean than in the Indian or Atlantic Oceans. At 10~dbar in the Pacific Ocean, there is again evidence of patterns that are likely to be related to Equatorial Rossby and Kelvin~waves.

The right panels in Figure~\ref{fig:corrLonLat} show the fitted correlations at the meridional lag $\Delta x_\mathrm{lat} = 800$~km (with $\Delta x_\mathrm{lon} = 0$, $\Delta t = 0$). The general conclusions are similar to above: the ranges are longer near the surface and in the Equatorial regions. The distinct patch of large correlations at 300~dbar in the Equatorial West Pacific is likely to be related to the Pacific thermocline which is tilted westward along the Equator. The meridional ranges also show interesting patterns in areas where the Amazon River and the Congo River flow into the Atlantic Ocean. Notice that all the plots in Figure~\ref{fig:corrLonLat} are on the same (logarithmic) color scale and can thus be compared directly. This comparison shows that the dependence structure is clearly anisotropic with the zonal ranges longer than the meridional~ones.

\begin{figure}[t]
	\centering
	
	\subfigure[300 dbar]{
		\includegraphics[width=8.5cm,trim = 5cm 0cm 0cm 6cm, clip=true]{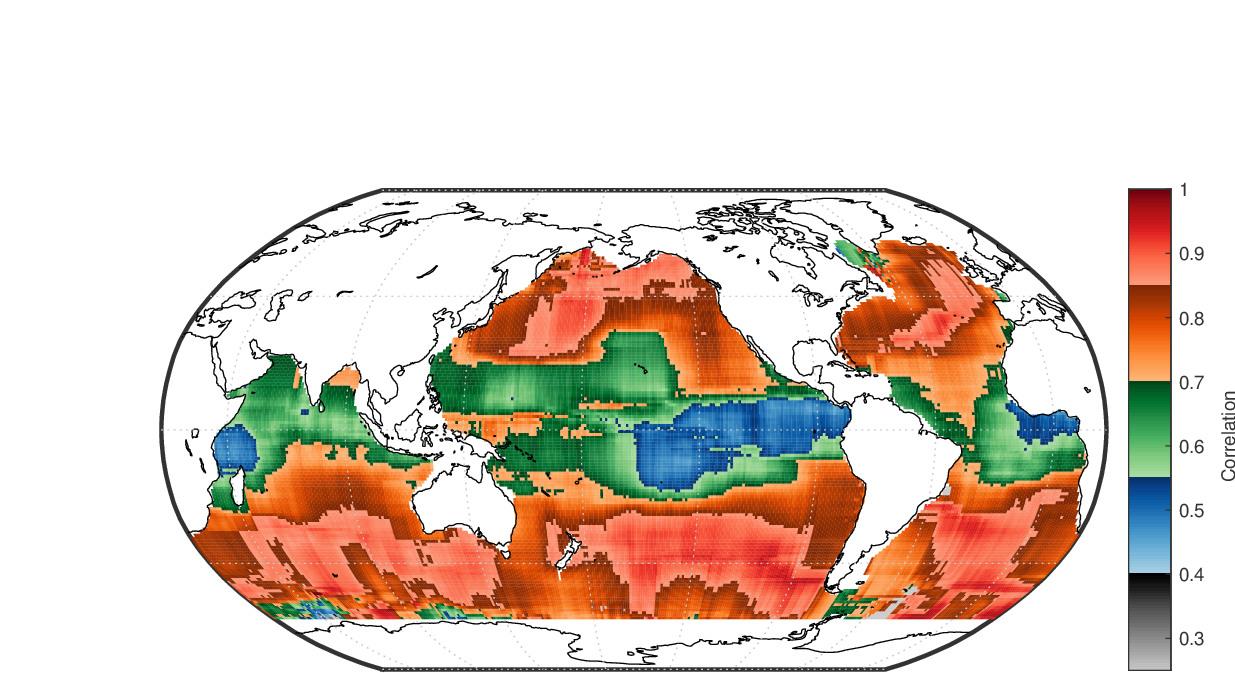}}\\
	\subfigure[10 dbar]{
		\includegraphics[width=7.7cm,trim = 5cm 0cm 0cm 6cm, clip=true]{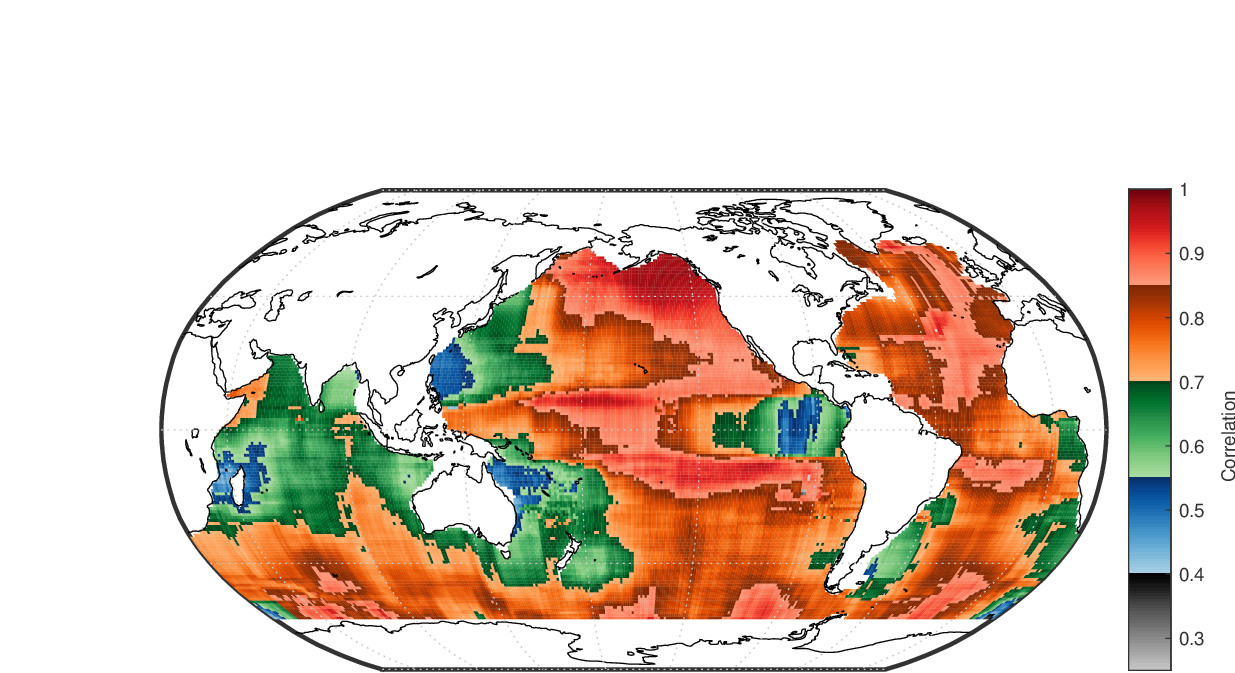}}
	\subfigure[1500 dbar]{
		\includegraphics[width=7.7cm,trim = 5cm 0cm 0cm 6cm, clip=true]{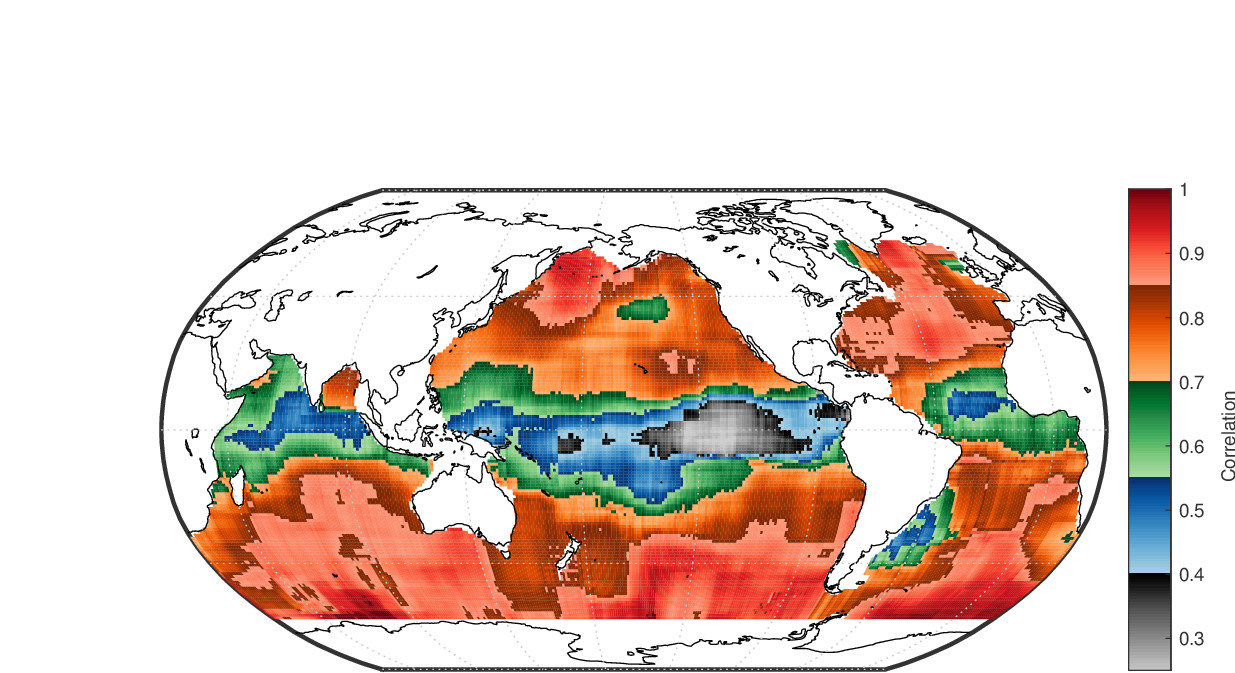}}
	
	\caption{Fitted correlations at temporal lag $\Delta t = 10$~days for the locally stationary 3-month spatio-temporal model with a Gaussian nugget (Model~5).}
	\label{fig:corrT}
\end{figure}

The fitted correlations at temporal lag $\Delta t = 10$ days (with $\Delta x_\mathrm{lat} = 0$ and $\Delta x_\mathrm{lon} = 0$) are given in Figure~\ref{fig:corrT}. While there are differences in the correlation patterns between the three pressure levels, the overall magnitude of the temporal correlation changes relatively little with pressure. There does seem to be a slight tendency for the temporal correlation to increase with pressure in the Southern Ocean and to decrease with pressure in most other areas, but these changes are much less pronounced than in the case of the zonal and meridional ranges (Figure~\ref{fig:corrLonLat}).

Ninove et al. \cite{Ninove2016} have previously estimated dependence scales using Argo data. They estimate the range parameters (which are often called ``decorrelation scales'' by oceanographers) by dividing the global ocean into large disjoint boxes and then use a weighted least-squares variogram fit to data within each box. They only consider spatial dependence and do not provide estimates of the temporal scales. In comparison, our moving-window approach includes the temporal dimension and provides information about the dependence structure at much finer horizontal resolution. It is also well-established in the spatial statistics literature that MLE fits should be preferred over variogram fits \cite{Stein1999}. Nevertheless, many of our conclusions qualitatively agree with those of Ninove et al. They also find longer ranges near the surface, zonal elongation in the tropics, strong anisotropy and shorter ranges in the Indian and Atlantic Oceans. However, we do not find evidence for the large increase in spatial ranges below 700~m that Ninove et al. observe. Instead, our spatial ranges are almost always shorter at 1500~dbar than at 300~dbar. We suspect that the effect seen by Ninove et al. is an artifact caused by the pre-Argo mean field that they use. This mean field is likely to be poorly constrained at depth, where little data was available before Argo, and any unmodeled nonstationary mean effects would then show up as increased spatial dependence in the empirical~variograms.

\section{Conclusions and outlook} \label{sec:discussion}

We have demonstrated that the spatio-temporal dependence structure of ocean temperatures can be estimated from Argo data using local moving-window maximum likelihood estimates of the covariance parameters. The resulting fully data-driven nonstationary anomaly fields and their uncertainties yield substantial improvements over existing state-of-the-art methods. The improvements in point prediction accuracy are comparable to deploying hundreds of new Argo floats. There is also evidence of non-Gaussian heavy tails in the temperature data and taking this into account is crucial for obtaining well-calibrated uncertainties. The estimated covariance models exhibit physically sensible patterns that can be analyzed further to test theories of large-scale ocean circulation.

The choice of the covariance parameters has been a long-standing conundrum in Argo mapping. The values of these parameters affect in particular how small-scale features, such as mesoscale eddies, are displayed on the map. It is sometimes argued that the covariance parameters should be chosen so that eddies are smoothed out from the final map. We do not regard this as a good basis for producing a general-purpose Argo data product. Instead, we believe that the covariance function should ideally reflect all the ocean variability present in the observations, including eddies and other small-scale features. The resulting map captures as much of the physical variation as possible and can be customized \emph{post hoc} for various purposes by applying operators on it. For example, an eddy-reduced map can be obtained by applying a low-pass filter on the map. This line of thinking allows the covariance function to model the actual ocean variability as well as possible, which is essentially what the data-driven length scales used in this paper are aiming to accomplish, and leaves the choice of which features to emphasize to the user of the data product, effectively decoupling these two issues.

Statistically, this work demonstrates how well-suited moving-window Gaussian process regression is for handling massive modern nonstationary spatio-temporal datasets. This approach helps address in a straightforward manner challenges related to both nonstationarity and computational complexity, issues that affect the analysis of almost any large-scale environmental dataset. In the present work, we developed a version of the moving-window approach that uses a Student-$t$ distributed nugget effect to address heavy tails in the temperature data. To the best of our knowledge, the Student nugget has not been previously used in conjunction with moving-window Gaussian process regression. While spatial and spatio-temporal moving-window techniques have been around since the seminal work of Haas \cite{Haas1990,Haas1995}, we feel that there is much room for further application and methodological development on this front.

To produce the interpolated temperature anomalies, we used the fairly simple exponential covariance function with geometric anisotropy along the zonal, meridional and temporal axes (see Equation~\eqref{eq:distanceMetric}). We have compared the model fit to empirical covariances in various regions and generally find the two to be in a reasonably good agreement. We have also investigated more complex space-time covariance models including the Mat\'{e}rn model~\cite{Stein1999}, the Gneiting model~\cite{Gneiting2002} and geometric anisotropy that is not constrained to be aligned with the latitude-longitude coordinate axes. We have experimented with these models in selected regions of the Pacific Ocean at 300~dbar. In each case, we found only minor improvements in the likelihood values and point predictions in comparison to the simple exponential covariance function. At the same time, the estimated covariance parameters became very challenging to interpret and validate since several combinations of the model parameters can represent almost the same covariance structure. Furthermore, the likelihood computations were much too slow to be practical on a global scale. These extensions could still be useful in other regions or pressure levels, but the interpretability and computations remain problematic. We have also experimented with various ways of adding the ocean depth into the distance metric in Equation~\eqref{eq:distanceMetric}, as is done for example in \cite{Roemmich2009}, but found that the corresponding range parameter is estimated to be very large in comparison to typical depth differences, which effectively removes this component from the model. We have also explored the possibility of accounting for land barriers using a simple line-of-sight approach, where those data points within the moving window that do not have a line-of-sight to the grid point at the center are discarded from the computations. We found that this way of handling land does not change the estimates much and can in fact lead to degraded performance since potentially useful data are left out from each prediction. An interesting future extension would be to add a velocity term to the covariance, as is for example done in the mapping of satellite altimetry data~\cite{Pujol2016}.

While the Student prediction intervals clearly yield improved uncertainty quantification, the present implementation of the Student nugget suffers from algorithmic instabilities, which is evident in the plots of the estimated model parameters given in the supplement \cite{Kuusela2017Supplement}. These instabilities are likely to be related to the nonconcave optimization problem that needs to be solved as part of the Laplace approximation \cite{Vanhatalo2009}. Alternatively, it might be that the Laplace approximation itself is not appropriate in some parts of the ocean. We experimented with an alternative implementation of the Laplace approximation \cite{Vanhatalo2013} but observed similar instabilities. Further research is needed to understand the precise cause of these instabilities before the Student models can be fully recommended for use in actual data products.

This work has only scratched the surface in terms of the statistical research that can be done with Argo data. We sketch below some potential extensions and directions for future work:
\begin{itemize}[leftmargin=0.7cm]
	\item In this work, we have focused on Argo temperature data only. In principle, similar locally stationary interpolation is also applicable to salinity data, but the challenge is that especially at larger depths the variability of the salinity field is so small that instrumental errors can no longer be ignored. This means that a more careful modeling of the nugget effect to include measurement error is needed. We also note that when mapping both temperature and salinity, one should ideally take into account their correlation which requires modeling the spatio-temporal cross-covariance between the two fields. This cross-covariance function can also be of scientific interest in its own right.
	\item Maps of the locally estimated covariance models (Figures~\ref{fig:thetasPlusSigma2}--\ref{fig:corrT}) tend to sometimes have zonal or meridional stripes that are related to the sharp boundary of the $20^\circ \times 20^\circ$ moving window. These stripes could be removed by replacing the moving window by a smooth kernel function as in \cite{Anderes2011}. Since it is often possible to represent almost the same covariance structure by various combinations of the model parameters, we also sometimes see abrupt transitions from one parameter configuration to another. It should be possible to fix this by finding a way to borrow strength across the nearby grid points in the parameter estimates. One could, for instance, envision using a Bayesian hierarchical model, where the maps of the local model parameters are themselves modeled as spatially dependent random fields, as in~\cite{Paciorek2006}, for example. However, it seems challenging to find a way to carry out the computations at the scale of the global ocean with such model.
	\item The type of non-Gaussianity considered in this work is spatially and temporally uncorrelated Student-$t$ distributed heavy tails. However, the remaining miscalibration in Figure~\ref{fig:qqPlot} indicates that either there is room for improvement in the covariance modeling or there is some non-Gaussian structure in the data that remains unaccounted for. Further exploratory analysis of the mean-subtracted residuals reveals for example the presence of skewness (especially near the surface, as can be expected based on previous analyses of sea surface temperature data \cite{Sura2008}) and regions where the distribution of the residuals appears to be multimodal. It is also highly likely that the heavy-tailed features, which at least partially correspond to eddies, are spatially and temporally correlated. Hence, an important area for future work would be to develop tools for understanding the spatial and temporal dependence scales of these heavy-tailed features. This would enable us to better understand how long these features persist and how large spatial regions are affected. Capturing all these effects would require developing space-time models with more versatile non-Gaussian structure. Stochastic partial differential equations, as in~\cite{Wallin2015}, may provide a way to construct such models.
	\item In the present work, we have not exploited dependence across pressure levels. A natural extension would be to carry out full 4D mapping by including the vertical dimension in the covariance structure. In principle, the moving-window approach can be easily extended to this situation by considering observations at nearby pressures. However, modeling the vertical covariance seems nontrivial since the vertical direction contains phenomena that are fundamentally different from the other dimensions. For example, ocean stratification can cause abrupt vertical changes, while eddies exhibit coherent vertical structure. One would also ideally like to find a way to incorporate the knowledge that in a stably stratified ocean, density needs to be a monotonically increasing function of pressure, so any violations of monotonicity must be constrained to be transient and small.
	\item Here we have only considered point-wise uncertainties. However, many scientifically important quantities are functionals of the temperature and salinity fields (or some field derived from these two quantities). For example, ocean heat content is essentially a 3D integral of the temperature field. Providing uncertainties for functionals requires access to predictive covariances or conditional simulations, but neither of these can be directly obtained from the moving-window approach considered here. It should, however, be possible to combine the local covariance models into a valid global nonstationary model using the approach developed in \cite{Paciorek2006}. This global model could then be used to compute the predictive covariance matrix or to produce conditional simulations, although it is not immediately clear if all the necessary computations are feasible on the scale of the Argo dataset.
\end{itemize}

\section*{Access to data, code and supplementary material}

The results presented in this paper are based on the May 8, 2017 snapshot of the Argo GDAC (\url{http://doi.org/10.17882/42182\#50059}). The Matlab code is available at \url{https://github.com/mkuusela/ArgoMappingPaper} and supplementary material can be found at \linebreak \url{https://github.com/mkuusela/ArgoMappingPaper/raw/master/Doc/supplement.pdf}.

\section*{Acknowledgments}

We are grateful to Fred Bingham, Chen Chen, Bruce Cornuelle, Donata Giglio, John Gilson, Sarah Gille, Alison Gray, Malte Jansen, Alice Marzocchi, Matt Mazloff, Breck Owens, Dean Roemmich, Megan Scanderbeg and Nathalie Zilberman for many discussions about Argo data, oceanography and previous mapping algorithms as well as for helpful feedback on preliminary results throughout this project. We would also like to thank the four anonymous referees and the editor for their detailed and insightful feedback which led to a much improved presentation of these results.

This work was carried out while MK was at the University of Chicago, Department of Statistics and at the Statistical and Applied Mathematical Sciences Institute. This work was supported by the STATMOS Research Network (NSF awards 1106862, 1106974 and 1107046), the US Department of Energy grant no. DE-SC0002557 and the National Science Foundation Grant DMS-1638521 to the Statistical and Applied Mathematical Sciences Institute. This work was completed in part using computational resources provided by the University of Chicago Research Computing Center.

Argo data were collected and made freely available by the International Argo Program and the national programs that contribute to it (\url{http://www.argo.ucsd.edu}, \url{http://argo.jcommops.org}). The Argo Program is part of the Global Ocean Observing System.

\section*{Disclaimer}

Any opinions, findings and conclusions or recommendations expressed in this material are those of the authors and do not necessarily reflect the views of the National Science Foundation or the US Department of Energy.

\bibliographystyle{unsrt}
\bibliography{references}

\end{document}